\documentclass[conference, letterpaper]{IEEEtran}
\pdfoutput=1
\usepackage{amsmath}
\usepackage{amsthm}
\usepackage{amssymb}
\usepackage{amsfonts}
\usepackage{cite}
\usepackage{url}

\usepackage{scalerel}

\DeclareMathOperator*{\lorrsc}{\scaleto{\lor}{9pt}}
\DeclareMathOperator*{\sumsc}{\scalerel*{\sum}{\lorrsc}}

\usepackage{import}

\usepackage{siunitx}
\DeclareSIQualifier\peak{p}
\DeclareSIQualifier\peakpeak{pp}
\DeclareSIUnit{\pixel}{px}

\usepackage{xcolor}
\usepackage{graphicx}

\usepackage{tikz}
\usepackage{pgfplots}
\usepackage[nooldvoltagedirection]{circuitikz}
\pgfplotsset{compat=1.14}
\usetikzlibrary{calc}
\usetikzlibrary{circuits.ee.IEC}
\usetikzlibrary{positioning}

\usepackage{textcomp}

\usepackage{listings}
\usepackage{adjustbox}
\usepackage{lipsum}
\usepackage{xparse}

\usepackage{placeins} %

\usepackage{ifthen}
\newboolean{arxiv}
\setboolean{arxiv}{true}

\usepackage[printwatermark]{xwatermark}
\ifthenelse{\boolean{arxiv}}{
\newwatermark*[firstpage,color=black!80,angle=0,xpos=-50,ypos=-128, align=center,fontsize=6.8,fontseries=m]{%
	\xwmminipage[width=.3\linewidth,textalign=justified]{\textcopyright~2020 IEEE. Personal use of this material is permitted. Permission from IEEE must be obtained for all other uses, in any current or future media, including reprinting/republishing this material for advertising or promotional purposes, creating new collective works, for resale or redistribution to servers or lists, or reuse of any copyrighted component of this work in other works.}}
\newwatermark*[firstpage,color=black!80,angle=0,xpos=49,ypos=-123.5, align=center, fontsize=6.8,fontseries=m]{%
	\xwmminipage[width=.3\linewidth,textalign=justified]{\textit{Note:} This is the authors' version of the article accepted for publication at IEEE Symposium on Security and Privacy 2021.}}
\newwatermark*[pages=2-17,color=black!80,angle=0,xpos=50,ypos=-120,align=left,fontsize=6.8,fontseries=m]{\textcopyright~2020 IEEE}
}{}

\newsavebox{\fminipagebox}
\NewDocumentEnvironment{fminipage}{m O{\fboxsep}}
{\kern#2\begin{lrbox}{\fminipagebox}
		\begin{minipage}{#1}\ignorespaces}
		{\end{minipage}\end{lrbox}%
	\makebox[#1]{%
		\kern\dimexpr-\fboxsep-\fboxrule\relax
		\fbox{\usebox{\fminipagebox}}%
		\kern\dimexpr-\fboxsep-\fboxrule\relax
	}\kern#2
}

\DeclareMathAlphabet{\mathitbf}{OML}{cmm}{b}{it}
\DeclareMathAlphabet{\mathbfit}{OML}{cmm}{b}{it}

\ifCLASSOPTIONcompsoc
 \usepackage[caption=false,font=normalsize,labelfont=sf,textfont=sf]{subfig}
\else
 \usepackage[caption=false,font=footnotesize]{subfig}
\fi

\begin{document}
\title{Real-World Snapshots vs. Theory:\\Questioning the $t$-Probing Security Model}

\author{
	\IEEEauthorblockN{Thilo Krachenfels\IEEEauthorrefmark{1},
	Fatemeh Ganji\IEEEauthorrefmark{2}\textsuperscript{\textsection},
	Amir Moradi\IEEEauthorrefmark{3}, 
	Shahin Tajik\IEEEauthorrefmark{2}\textsuperscript{\textsection} and
	Jean-Pierre Seifert\IEEEauthorrefmark{1}}
	\IEEEauthorblockA{\IEEEauthorrefmark{1} Technische Universit\"at Berlin, Chair of Security in Telecommunications, Germany\\
	}
	\IEEEauthorblockA{\IEEEauthorrefmark{2} Worcester Polytechnic Institute, Department of Electrical and Computer Engineering, USA\\
	}
	\IEEEauthorblockA{\IEEEauthorrefmark{3} Ruhr-Universit\"at Bochum, Horst G\"ortz Institute for IT-Security, Germany\\
	}
}

\maketitle

\ifthenelse{\boolean{arxiv}}{
\thispagestyle{plain}
\pagestyle{plain}
}{}

\begingroup\renewcommand\thefootnote{\textsection}
\footnotetext{These authors contributed to this work when they were with Technische Universit\"at Berlin.}
\endgroup

\begin{abstract}
Due to its sound theoretical basis and practical efficiency, masking has become the most prominent countermeasure to protect cryptographic implementations against physical side-channel attacks~(SCAs).
The core idea of masking is to randomly split every sensitive intermediate variable during computation into at least $\mathbfit{t}\mathbf{+1}$ shares, where $\mathbfit{t}$ denotes the maximum number of shares that are allowed to be observed by an adversary without learning any sensitive information.
In other words, it is assumed that the adversary is bounded either by the possessed number of probes (e.g., microprobe needles) or by the order of statistical analyses while conducting higher-order SCA attacks (e.g., differential power analysis).
Such bounded models are employed to prove the SCA security of the corresponding implementations. 
Consequently, it is believed that given a sufficiently large number of shares, the vast majority of known SCA attacks are mitigated.

In this work, we present a novel laser-assisted SCA technique, called Laser Logic State Imaging~(LLSI), which offers an unlimited number of \emph{contactless} probes, and therefore, violates the probing security model assumption.
This technique enables us to take \emph{snapshots} of hardware implementations, i.e., extract the logical state of all registers at any arbitrary clock cycle with a single measurement.
To validate this, we mount our attack on masked AES hardware implementations and practically demonstrate the extraction of the full-length key in two different scenarios.
First, we assume that the location of the registers (key and/or state) is known, and hence, their content can be directly read by a single snapshot.
Second, we consider an implementation with unknown register locations, where we make use of multiple snapshots and a SAT solver to reveal the secrets. 
\vspace{5pt}
\end{abstract}


\IEEEpeerreviewmaketitle

\setcounter{footnote}{0}

\section{Introduction}
\label{sec:introduction}
Electronic embedded devices are an indispensable part of our today's connected systems.
To ensure the confidentiality and integrity of processed data in these systems, strong cryptography is needed.
But even in the presence of such cryptographic primitives, the security of deployed devices still can be compromised by attackers, who can gain access to these devices and thus launch physical attacks.
Side-Channel Analysis~(SCA) attacks are examples of such physical threats, which are hard to detect and mitigate due to their most often passive nature.
SCA attacks exploit the inevitable influence of computation and storage on different measurable quantities on a device, such as timing~\cite{kocher1996timing}, power consumption~\cite{kocher1999differential}, ElectroMagnetic~(EM) emanation~\cite{agrawal2002side}, and photon emission~\cite{ferrigno2008aes}.

Several countermeasures have been proposed to defeat SCA attacks.
Among them, masking has been shown to be the most effective one that can be applied to most cryptographic schemes.
Masking schemes are based on the principle of splitting the computation over several randomized and independent shares.
To prove the security of the masked implementations, the $t$-probing model was first introduced in the seminal work of Ishai et al.~\cite{ishai_private_2003}.
In this model, the adversary is assumed to be limited by the number of $t$ probes available for observing the computation on wires.
In such a scenario, we require to employ at least $t+1$ shares to assure that the adversary cannot learn any sensitive information from $t$ observations. 
In practice, assuming such a limit is quite plausible.

For instance, due to the lack of spatial distance in case of invasive micro/nano-probing attacks or EM analysis, we expect the number of possible probes to be very limited.
Moreover, the higher number of probes leads to a more expensive probe station, and hence, the cost of multi-probe stations is another limiting factor for the adversary.
Currently, the most advanced commercially-available nano-probe station consists of at most eight needles~\cite{kleindiek}.
Similarly for EM stations, the largest setup, which has been reported so far only in~\cite{specht_dividing_2018}, makes use of three simultaneous probes.
In the case of classical power analysis, typically only one physical probe is available.
However, it captures the entire circuit's power consumption, including that of all shares of all sensitive variables at once (univariate) or at multiple time instances (multivariate).
Therefore, higher-order statistical analyses dealing with such power measurements to some extent reflect the number of probes, for example, see~\cite{DBLP:journals/joc/DucDF19}.
Such higher-order analyses are, however, strongly affected by the noise level~\cite{DBLP:journals/tc/ProuffRB09}.
Consequently, it is believed that employing a sufficiently large number of shares can -- in the presence of noise -- avert classical SCA attacks.

On the other hand, more advanced photonic SCA attacks from the chip backside~\cite{carmon2017photonic} enable the adversary to capture side-channel information of several transistors simultaneously, and hence, can provide a large number of probes.
However, these attacks can only extract data during transitions.
Moreover, due to the typically low Signal-to-Noise Ratio~(SNR), the integration of leakages associated to many executions of the cryptographic algorithm with attacker-controlled inputs is necessary.
Yet, the existing randomization in masking schemes makes measurement repetition and integration over the same data infeasible.
While randomization has been mainly considered as a countermeasure against power/EM SCA attacks in the literature, optical attacks become ineffective as well due to their need for integration.

In response, an intriguing research direction dealing with single-trace SCA attacks has been formed, which mainly target the implementation of public-key algorithms requiring a large number of clock cycles~\cite{primas_singletrace_2017, jarvinen_singletrace_2017, alam2018one}.
Besides, there have been efforts to mount SCA with a minimum possible number of traces by profiling the target in advance, also known as template attacks~\cite{lerman2015template,specht_dividing_2018}.
Unfortunately, these techniques are relevant only for specific cryptographic schemes and cannot be applied in general to all masked implementations.
Furthermore, the profiling phase, in the case of template attacks, might be infeasible in real-world scenarios, where only one sample is available.
Besides, it should be noted that profiling still does not guarantee the success of the SCA attack by a single-trace measurement and cannot easily scale with an increase in the number of shares.
Driven by the limitations mentioned above, the following question arises: \textit{Does a practical single-trace SCA technique exist that offers an unlimited number of probes while not being limited to specific cryptographic algorithms?}

\noindent
\textbf{Our Contributions.} In this work, we indeed positively answer the above question. 
We present a novel laser-assisted SCA attack from the chip backside using a known Failure Analysis~(FA) technique, called Laser Logic State Imaging~(LLSI)\footnote{It should be noted that conducting LLSI from the IC backside has been previously reported in the failure analysis community. We claim neither this technique nor our experimental setup as the contribution of this work. Our primary intention is to draw attention to the potential threat of this known but not well-researched technique as an attack tool.}.
By modulating the voltage supplying the transistors on the chip, the corresponding light reflection (originating from a laser scanning irradiation on these transistors) also becomes modulated.
The resulting modulation is highly data dependent because only transistors in the on-state affect the reflection of the laser.
We demonstrate how an adversary can deploy LLSI in a particular clock cycle to take a \emph{snapshot} from the entire circuit and recover the state of all transistors, which form the gates and registers.
Hence, it enables the adversary to have an unlimited number of \emph{contactless} probes during a time period, which invalidates the central underlying assumption of the probing security model for masking schemes.
Moreover, in contrast to other optical attacks or conventional SCA techniques, LLSI does not require any repeated measurements with the same data.
Therefore, the existing randomness in masking schemes does not have any protective effect.

To validate our claims, we consider two attack scenarios.
First, we assume that the location of the registers is known to the adversary; hence their content can be directly read out using a single snapshot.
If this includes key and/or state registers of the underlying cipher, extracting the secret key is straightforward.
In this case, the effort for the attacker grows linearly with the number of shares.
Second, we demonstrate that even without knowing the location of the registers, the attacker can still recover the secret key by 
capturing a couple of snapshots at consecutive clock cycles, and making use of a SAT solver.
Apart from several simulation-based investigations, to practically show the effectiveness of LLSI we mount snapshot attacks on masked AES designs implemented on a Field Programmable Gate Array~(FPGA) manufactured with a 60\,nm technology.
As a result, we successfully break the security of the targeted masked implementations by extracting their full-length keys.

\section{Background}
\label{sec:background}

\begin{figure*}
	\centering
	\subfloat[Classical EOFM
	]{
		{\small
			\def\svgwidth{.45\textwidth}
			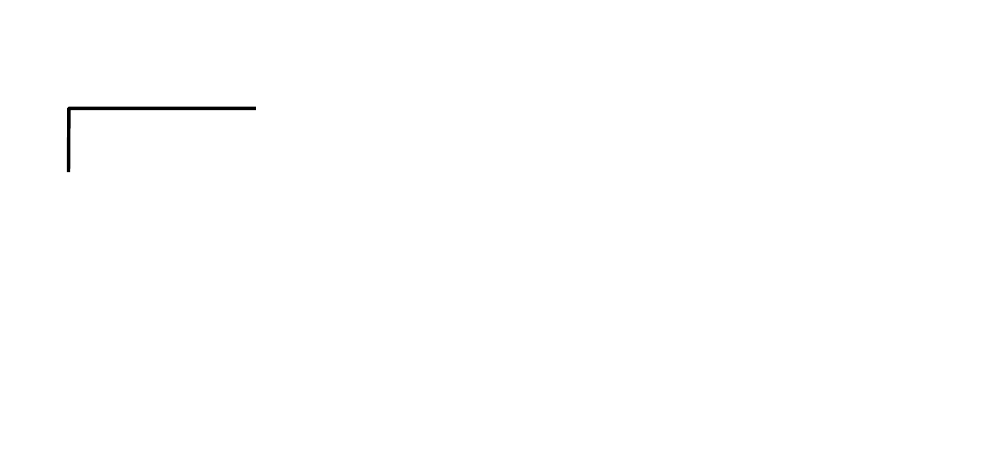
		}
		\label{fig:background_eofm}}
	\hfil
	\subfloat[LLSI
	]{
		{\small
			\def\svgwidth{.45\textwidth}
			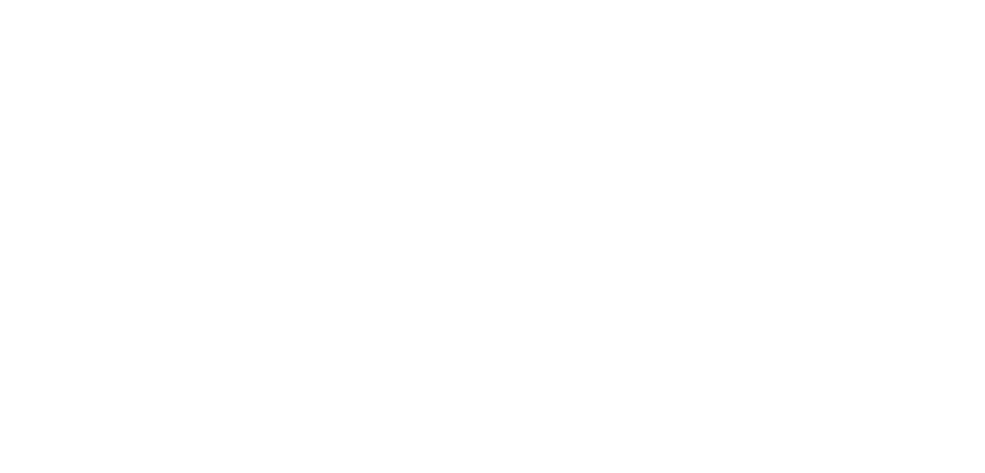
		}
		\label{fig:background_llsi}}
	\setlength{\abovecaptionskip}{4pt}
	\setlength{\belowcaptionskip}{-4pt} %
	\caption{Comparison of classical EOFM with LLSI.
		Classical EOFM can be applied to localize transistors switching with a known data-dependent frequency (here: 1\,\si{\mega\hertz}), however, transistors carrying a static signal do not appear in the image.
		In contrast, for LLSI, the power supply is modulated with a known frequency (here: 2\,\si{\mega\hertz}), thus transistors in the on-state can be localized.}
	\label{fig:background_eofmLlsi}
\end{figure*}

\subsection{Masking Countermeasures and \texorpdfstring{$t$}{t}-Probing Model}
\label{sec:background_masking}
While several customized countermeasures (e.g., shielded hardware, current filtering, and dual-rail logic) have been designed to protect specific cryptographic implementations against SCA attacks, masking is known as the most widely studied one with sound theoretical and mathematical foundations.
The main idea behind masking schemes is to make use of a couple of parties (order of the masking), and split the intermediate computations dealing with the secrets, i.e., multi-party computation and secret sharing.
The input of the circuit (key and plaintext) should be represented in a shared form, and the final result (ciphertext) should be obtained by recombining the output shares while the entire computations are performed only on shares. 
The primary advantage of masking is that it can be assessed in formal security models.
In Boolean masking, as the most common scheme, every random bit $x$ is represented by $(x_0,\ldots,x_d)$ in such a way that $x=x_0 \oplus \ldots \oplus x_d$.
Based on formal analyses given in~\cite{chari1999towards}, a secret sharing with $d+1$ shares can at most defeat an adversary who is limited to the $d$\textsuperscript{th} order SCA. 
Further, it has been demonstrated that measurements of each share $x_i$ are affected by Gaussian noise, and hence, the number of noisy traces required to recover $x$ grows exponentially with the number of shares~\cite{DBLP:journals/tc/ProuffRB09}.
Therefore, as a general knowledge, a higher number of shares would potentially diminish the feasibility of attacks. 

On the other hand, the security of masking has been analyzed by the $t$-probing model, which was first introduced in~\cite{ishai_private_2003}.
In this model, it is assumed that the adversary has access to at most $t$ physical probes to observe the computation on wires of the circuit at each time period (e.g., one clock cycle).
In such a scenario, at least $t+1$ shares are required to ensure that the adversary cannot learn any sensitive information from $t$ observations.
Although we would like to consider an adversary with an unlimited number of probes, this task is generally impractical according to the impossibility of obfuscation~\cite{ishai_private_2003, barak2001possibility}.
To unify the leakage models, and therefore, simplify the analysis of SCA countermeasures, it has been shown that the two aforementioned leakage models are related by reducing the security in one model to the security of the other one~\cite{DBLP:conf/eurocrypt/DucDF14,DBLP:journals/joc/DucDF19}.
In other words, a $d$\textsuperscript{th}-order noisy SCA is equivalent to placing $t=d$ physical probes on the wires of the target circuit.
Based on such models and assumptions, several constructions have been introduced~\cite{DBLP:journals/joc/NikovaRS11,DBLP:conf/asiacrypt/BilginGNNR14,DBLP:conf/crypto/ReparazBNGV15,DBLP:conf/eurocrypt/BalaschFG15,DBLP:journals/jce/GrossM18,DBLP:journals/tches/GrossIB18}, and a couple of security proofs have been given~\cite{DBLP:conf/eurocrypt/DucFS15,DBLP:conf/eurocrypt/DziembowskiFS15,DBLP:conf/eurocrypt/BartheDFGSS17,DBLP:conf/crypto/BronchainHMOS19}. 
Moroever, some (security) verification tools have been developed~\cite{DBLP:conf/eurocrypt/BartheBDFGS15,DBLP:conf/ccs/BartheBDFGSZ16, DBLP:conf/eurocrypt/BloemGIKMW18,DBLP:conf/icecsys/ArribasNR18,DBLP:conf/esorics/BartheBCFGS19}, and multiple implementations have been reported~\cite{DBLP:conf/ches/CnuddeRBNNR16,DBLP:journals/tcad/BilginGNNR15,DBLP:conf/cosade/Wegener018,DBLP:journals/tches/Meyer0W18,DBLP:journals/joc/PoschmannMKLWL11,DBLP:conf/eurocrypt/MoradiPLPW11,DBLP:conf/ctrsa/GrossMK17}.

In order to highlight the deployment of masking schemes in real-world products, we would like to mention that protection against side-channel attacks is among the criteria defined by certification bodies in several countries.
Masking schemes are among the countermeasures which have been employed in, e.g., banking cards since more than a decade ago by smartcard vendors.

\subsection{Optical Backside Failure Analysis Techniques} %
\label{sec:background_optical}

Due to the increasing number of metal layers on the frontside of integrated circuits (ICs), optical FA techniques have been developed to access on-chip signals through the backside~\cite{boit2016ic}.
The main techniques are photon emission analysis, laser stimulation, and optical probing, which take advantage of the high infrared transmission of silicon for wavelengths above 1\,\si{\micro\meter}.
Although initially developed for FA purposes, these techniques are nowadays also used in the security domain~\cite{rahman_physical_2018}.
FA labs are equipped with machines that incorporate all of the previously mentioned techniques in one device, which is typically a laser scanning microscope (LSM) equipped with a camera for photon emission analysis, a detector for measuring the reflected laser light, and laser sources of different wavelengths.\footnote{For a discussion on cost and availability of such FA machines, see Section~\ref{sec:discussion_cost}.}

Due to their high spatial resolution, optical FA techniques seem to be promising for conducting single-trace measurements.
For instance, the analysis of Photon Emission (PE) with temporal resolution allows to detect the time of switching activities of single transistors.
Related techniques are Picosecond Imaging Circuit Analysis~(PICA)~\cite{mcmanus_pica_2000}, and the more low-cost approach of Simple Photonic Emission Analysis (SPEA)~\cite{schlosser_simple_2012}, which has been used to attack unprotected implementations of, e.g., AES~\cite{schlosser_simple_2012} and RSA~\cite{carmon2017photonic}.
However, the circuit has to be repeatedly stimulated for these techniques, since the emission probability is very low for a single switching event. 
This disqualifies time-resolved PE analysis from being a single-trace technique.
Optical techniques that in principle can probe static signals are Thermal Laser Stimulation~(TLS)~\cite{lohrke_key_2018} and spatial PE analysis of off-state leakage current~\cite{stellari_revealing_2016, couch_direct_2018}.
However, due to the requirements of low noise on the power line for TLS, and high static current for PE analysis, these techniques are restricted to specific applications and targets.
In contrast, optical probing seems to be a more promising technique, and thus, it is discussed in more detail below.

\subsubsection{Optical Probing - EOP and EOFM} %
For optical probing, a laser beam is focused by a microscope-based setup on the backside of the IC, and the reflected light is analyzed to find data dependencies.
Since the refractive index and absorption coefficient within the silicon depend on the electrical properties present in the device~\cite{kindereit_quantitative_2007}, the laser light irradiating the IC is modulated and partially reflected.
A detector processes the returning light and converts it to an electrical signal.
Due to the transparency of the silicon to the wavelengths above 1.1\,\si{\micro\meter}, optical probing can be carried out in a non-invasive manner on some devices~\cite{tajik_power_2017} (see also Section \ref{sec:discussion_preparation}).

The laser can either be parked at a specific location, or scanned over a larger area of the chip.
When the laser remains at a particular location, the waveform of the signal of interest can be extracted.
This technique is called Electro-Optical Probing~(EOP)\footnote{When using a coherent light source, EOP is typically called Laser Voltage Probing~(LVP), and EOFM is called Laser Voltage Imaging~(LVI).}.
\newcounter{fnnumber}
\setcounter{fnnumber}{\thefootnote}
To achieve a sufficiently-high SNR, many repetitions of the same waveform must be integrated.
On the other hand, when the laser scans an area, the detected signal can be fed into a spectrum analyzer set to a narrowband filter for finding areas on the chip that operate with a specific frequency.
This technique is known as Electro-Optical Frequency Mapping~(EOFM)\footnotemark[\thefnnumber].
The result of an EOFM measurement is a 2-D image showing a signature at areas switching with the frequency of interest, see Fig.~\ref{fig:background_eofm}.

Two crucial steps are involved in an attack scenario where the adversary tries to localize and probe a set of re\-gis\-ters/me\-mo\-ries using optical probing~\cite{lohrke_no_2016, tajik_power_2017}.
First, the attacker induces a known frequency into the device (e.g., by supplying the clock or rebooting the chip at a specific frequency) to activate the target registers or memories, see Fig.~\ref{fig:background_eofm}.
Second, the device is operated in a loop, and EOP can be used to read out the values of each individual register.
Note that if the sensitive data are processed in parallel, the content of the registers can be directly obtained from the EOFM image~\cite{lohrke_no_2016}.
As a result, EOFM can be deployed to localize and probe the secret simultaneously on a cryptographic device.
However, the downside of this approach is that only dynamic signals which are available for an arbitrary number of repetitions can be extracted.
Therefore, classical EOP/EOFM cannot be used to extract static data, i.e., the state of memory elements that are only available once and at a certain point in time.

\subsubsection{LLSI}
Laser Logic State Imaging (LLSI) makes the readout of static signals possible. 
The technique was introduced as an extension to EOFM to the failure analysis community~\cite{niu_laser_2014}.
For LLSI the supply voltage is modulated with a known frequency.
Due to the modulation of the transistor channel's electric field caused by the supply voltage modulation, transistors in the on-state give clear signatures on the LLSI image, while this is not the case for transistors in the off-state, see Fig.~\ref{fig:background_llsi}.
This observation can be used to deduce the logical state of, for instance, a memory cell.

\begin{figure}
	\centering
	\begin{tikzpicture}[font=\small,circuit ee IEC, scale=0.85]
	\begin{scope}
	\path[draw] (0, 0) node[nmos, xscale=-1] (n1) {};
	\path[draw] (n1.drain) -- node[contact, pos=.25] (out1) {} ($(n1.drain)+(0,1)$) node[pmos, anchor=source, yscale=-1, xscale=-1] (p1) {};
	\path[draw] (n1.gate) -- (p1.gate);
	
	\path[draw] (out1) -- ($(out1)+(1.5, 0)$) node[contact] (in2) {};
	\path[draw] (in2) -- (in2 |- p1.gate) node[pmos, anchor=gate, yscale=-1] (p2) {};
	\path[draw] (in2) -- (in2 |- n1.gate) node[nmos, anchor=gate] (n2) {};
	\path[draw] (p2.source) -- node[contact, pos=.25] (out2) {} (n2.drain);
	
	\path[draw] (n1.gate |- out2) node[contact] (in1) {} -- (out2);
	
	\path[draw] (n1.source) -- ($(n1.source)+(0,-0.25)$) coordinate (gnd);
	\path[draw] (n2.source) -- (n2.source |- gnd) -- node[contact, midway] (gnd) {} (gnd);
	\path[draw] (p1.drain) -- ($(p1.drain)+(0,0.25)$) coordinate (vcc);
	\path[draw] (p2.drain) -- (p2.drain |- vcc) -- node[contact, midway] (vcc) {} (vcc);
	
	\path[draw] (gnd) -- ($(gnd)+(0,-0.25)$) node[ground, anchor=input, rotate=-90] {} node[anchor=north,yshift=-9] {GND};
	\path[draw] (vcc) to[short,-o] ($(vcc)+(0,0.25)$) node[anchor=south] {VCC};
	\begin{scope}[scale=0.3]
	\draw[x=1ex, y=1ex, thick, red, xshift=170, yshift=410] (3,0) sin (4,2) cos (5,0) sin (6,-2) cos (7,0) sin (8,2) cos (9,0) sin (10,-2) cos (11,0);
	\end{scope}
	\begin{scope}[scale=0.3]
	\draw[x=1ex, y=1ex, thick, red, xshift=170, yshift=-185)] (3,0) sin (4,2) cos (5,0) sin (6,-2) cos (7,0) sin (8,2) cos (9,0) sin (10,-2) cos (11,0);
	\end{scope}  
	
	\path[fill=black!50] ($(p2.drain)+(-.15,-0.55)$) rectangle ($(p2.source)+(0,+0.55)$);
	\path[fill=black!50] ($(n1.drain)+(+.15,-0.55)$) rectangle ($(n1.source)+(0,+0.55)$);

	\node[draw=red, very thick, shape=rectangle, anchor=center, minimum width=1cm, minimum height=1cm] (rectlegend) at ($(p2.gate)+(.65,0)$) {};	
	\node[draw=red, very thick, shape=rectangle, anchor=center, minimum width=1cm, minimum height=1cm] (rectlegend) at ($(n1.gate)+(-.65,0)$) {};
	
	\node[left=1.2cm of p1.gate] {P1};
	\node[left=1.2cm of n1.gate] {N1};
	\node[right=1.2cm of p2.gate] {P2};
	\node[right=1.2cm of n2.gate] {N2};
	
	\path ($(vcc)+(3,0)$) coordinate(legendbase); 
	\begin{scope}[shift={(legendbase)}]%
	\path[draw, black, line width=3pt, opacity=0.3]
	(-0.15, -0.15) -- (0.15, 0.15)
	(0.15, -0.15) -- (-0.15, 0.15);
	\path[draw] (0,-0.04) node[anchor=west,xshift=4] {High-Ohmic Channel};
	\end{scope}
	\begin{scope}[shift={($(legendbase)+(0,-0.5)$)}]%
	\path[draw, black, line width=3pt, opacity=0.5]
	(0, -0.15) -- (0, 0.15);
	\path[draw] (0,-0.00) node[anchor=west,xshift=4] {Low-Ohmic Channel};
	\end{scope}
	\node[draw=red, thick, shape=rectangle, anchor=center] (rectlegend) at ($(legendbase)+(0,-1)$) {};
	\path[draw] (rectlegend) node[anchor=west,xshift=4] {LLSI Modulation Areas};
	
	\path ($(gnd)+(3.8,0)$) coordinate(blockbase);       
	\path[fill=black!75] ($(blockbase)-(0,0.3)$) rectangle ($(blockbase)+(2,2.3)$);
	\path[fill=black!10] ($(blockbase)+(.3,.3)-(0,.3)$) rectangle ($(blockbase)+(.9,.9)-(0,.3)$);
	\path[fill=black!10] ($(blockbase)+(1.1,1.1)+(0,.3)$) rectangle ($(blockbase)+(1.7,1.7)+(0,.3)$);
	\draw ($(blockbase)+(1,2.3)$) node[anchor=south] {Simplified LLSI Image:};
	
	\foreach \t in {n2, p1} {
		\begin{scope}[shift={(\t)}]
		\path[draw, black, line width=5pt, opacity=0.3]
		(-0.25, -0.25) -- (0.25, 0.25)
		(0.25, -0.25) -- (-0.25, 0.25);
		\end{scope}
	}
	
	\end{scope}
\end{tikzpicture}
	\setlength{\abovecaptionskip}{-2pt}
	\setlength{\belowcaptionskip}{-10pt} %
	\caption{Schematic of a CMOS memory cell and the expected 2-D LLSI image for the cell. For simplicity we omit the input transistors. Only the transistors in the on-state are expected to give a strong LLSI signal, therefore, the logic state of the memory cell can be deduced. Figure based on~\cite{niu_laser_2014}.\vspace{-10pt}} \label{fig:background_llsiSRAM}
\end{figure}
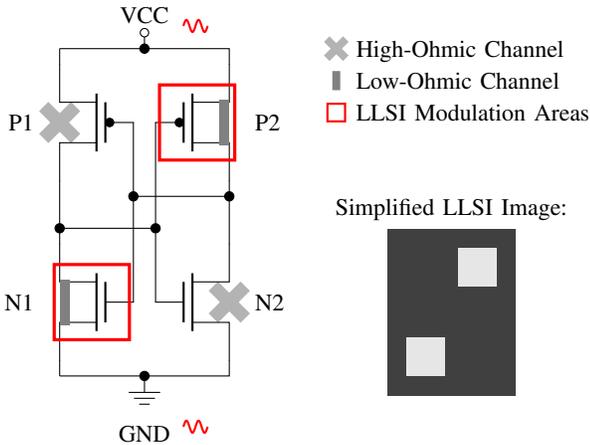

\begin{figure*}
	\centering
	{
		\def\svgwidth{.80\linewidth}
		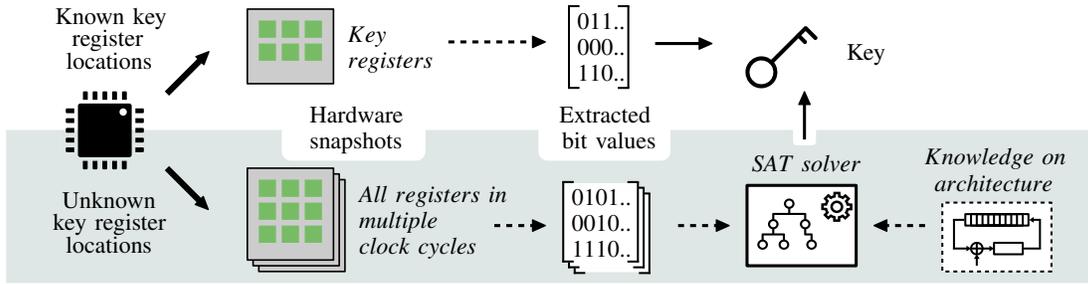
	}
	\setlength{\abovecaptionskip}{0pt}
	\caption{Two approaches with different assumptions: known key register locations (Section~\ref{sec:approach_direct}) and unknown key register locations (Section~\ref{sec:approach_localization}).\vspace{-10pt}}
	\label{fig:scenario_approach}
\end{figure*}

Fig.~\ref{fig:background_llsiSRAM} shows a CMOS memory cell consisting of two cross-coupled inverters.
Each inverter consists of one PMOS and one NMOS transistor, connected between VCC and GND.
The input to the CMOS inverter directly dictates whether its NMOS is in the on-state and the PMOS transistor in the off-state, or vice-versa.
In both cases, only one transistor per inverter is in the on-state.
Consequently, when knowing the transistors' states, the value of the inverters' input can be derived.
By modulating the power supply of the device, the channel's electric field of all transistors in the on-state modulates with the induced frequency and, as explained above, that can be detected using LLSI.
In the example given in Fig.~\ref{fig:background_llsiSRAM}, the top right and bottom left transistors are in the on-state, and the expected simplified LLSI image shows a clear signature at those two locations.
With the inverted input values, the other two transistors would be in the on-state, resulting in clear signatures on the top left and bottom right of the image.
Hence, it can be concluded that all logic states can be extracted using one LLSI measurement.
\section{Threat Model}\label{sec:scenario}
With our attack, we target hardware implementations of a block cipher protected by some masking countermeasure.
While assuming here that the input (plaintext/ciphertext) and the key are shared by Boolean masking, we do not presume any specific masking scheme.
Note that the key has to be stored in a masked format on the chip, and it has to be re-masked with fresh masks every time it is used.
Otherwise, template attacks~\cite{wagner2017brute} or classical optical probing~\cite{lohrke_no_2016} on key or key schedule might be possible.
The cipher might be implemented on an FPGA or realized as an ASIC.
Following the common serialized or round-based design architecture, or as being enforced by the glitch-resilient masking schemes, the implementation should make use of registers to store the cipher's intermediate values.

We stress that in our technique, we are not making use of any specific construction or feature of any certain masking scheme. 
We just suppose that the state register (and key register) are masked, which is a general statement and does not deal with any particular technique to realize masking schemes in hardware, like TI~\cite{DBLP:journals/tcad/BilginGNNR15}, DOM~\cite{gross_domainoriented_2016}, GLM~\cite{DBLP:journals/tches/GrossIB18}, CMS~\cite{DBLP:conf/crypto/ReparazBNGV15}, UMA~\cite{DBLP:journals/jce/GrossM18}, etc.
Note that these different masking schemes define various techniques to realize non-linear functions (like the ciphers' Sboxes), but they all have in common that the state and key registers are masked. 
In short, even if the underlying Boolean masking scheme of the target device does not follow any of the known hardware masking schemes, our approach is still a valid attack vector.

Under the above assumptions, we consider a potential attacker, who can take snapshots of the hardware state using LLSI and extract the values stored in the registers.
To read out the content of registers at a specific clock cycle, the attacker should either halt the clock or the content should remain in the registers and not get cleared after the algorithm has terminated (see Section~\ref{sec:discussion_clock} for a detailed discussion on clock control).
For the purpose of extracting the secret, the attacker could either directly target the (masked) key registers or some registers containing intermediate values of the cipher, from which the secret can be deduced.
Which registers the attacker chooses to target, depends on her knowledge about the netlist and layout of the implementation.
Regarding this, two scenarios can be discussed (see Fig.~\ref{fig:scenario_approach}).
\textbf{Scenario~1}: If the attacker knows where the key registers are located on the chip, possibly learned by reverse engineering, she could directly target them.
Still, due to the underlying masking scheme, she has to target all shares of the key registers.
We consider this as the most straightforward scenario and cover it in Section~\ref{sec:approach_direct}.
\textbf{Scenario~2}: If the attacker does not know which registers on the chip contain the secret, some knowledge about the algorithm can help with the key extraction, as explained in Section~\ref{sec:approach_localization}. 
Related to this, we also propose a method to differentiate registers from other combinatorial gates on a chip, if the attacker does not even know the areas on which the registers of the design are placed.

\noindent
\textbf{\textit{Real-World Targets.}}
To demonstrate how an adversary might benefit from such an attack in the real world, we provide some examples for the target devices.
One example would be payTV smartcards, which are all programmed with the same key to decrypt the scrambled satellite signal in the receivers using some block cipher.
By extracting the encryption key, the adversary can counterfeit the payTV cards and sell them in the black market.
Consequently, extracting the secret from one device breaks the security of all devices in the field.
Another example would be every microcontroller/microprocessor or FPGA that supports firmware or bitstream encryption, respectively.
If the adversary can break this protection mechanism by extracting the key, she can decrypt the firmware/bitstream and clone, reverse-engineer, or tamper with the IP.
Note that the adversary is not interested in the hardware itself, and hence, even if the chip gets unusable during the key extraction, the main assets, e.g., key or firmware, are still valuable for the adversary.

\section{Approach}\label{sec:approach}
This section describes methods employed to launch our attack in Scenario~1 and Scenario~2 explained in Section~\ref{sec:scenario}. 

\vspace{-1mm}
\subsection{Scenario~1: Known Register Locations}\label{sec:approach_direct}
Here we assume that the location of the key registers (i.e., registers used to store key shares) on the chip is known to the adversary.
In this case, at some point in time, a given secret key (in a shared form) is loaded in these key registers. 
Once the attacker knows the corresponding clock cycle, she can take snapshots of the chip using LLSI.
The attacker, in principle, can learn the location of these registers by reverse-engineering the layout and netlist of the chip.
In the case of an ASIC, this can be done by de-layering the chip and applying some tools to extract the netlist (e.g., ChipJuice ~\cite{texplained_hardware_2018}).
Interestingly enough, the whole procedure is also available as a service, e.g.,~\cite{techinsigh_access_2020}.
If the implementation platform is an FPGA, reverse-engineering the netlist from the bitstream is essential~\cite{pham2017bitman, zhang_comprehensive_2019, ender2019insights}.
When the bitstream is available solely in an encrypted form, the attacker first needs to decrypt it. 
This is possible, as most cryptographic ASIC cores on mainstream SRAM-based FPGAs, responsible for decrypting the bitstream, are either not protected against SCA or contain other implementation vulnerabilities~\cite{moradi2011vulnerability, moradi2013side, moradiimproved, tajik_power_2017, lohrke_key_2018, ender2020unpatchable}. 
Moreover, it is worth mentioning that an attacker, who is involved in the development and fabrication process of the IC or  has enough influence on those entities, might possess parts or entire information necessary to localize the (key) registers on the chip.

\noindent
\textbf{\textit{Automatically extracting bit values from snapshots.}}
To extract the values from the register snapshots, the attacker first has to discover the data dependency in the LLSI measurements.
To this end, if she has control over the data written in the registers, she can take two snapshots of a register cell containing once the value \texttt{0} and another time~\texttt{1}.
By subtracting these LLSI images from each other, the attacker can clearly localize the data dependency.
Upon knowing how to distinguish between \texttt{0} and \texttt{1}, she can extract the values in an automated fashion.

For this purpose, we propose an approach based on classical image processing techniques, namely image registration through cross-correlation, cf.~\cite{sarvaiya2009image}.
For this, the corners of each register cell (containing one bit of data) should be known with sufficient precision so that the attacker can cut the snapshot of a single register cell from a potentially larger image.
For selecting the cell boundaries on an FPGA, domain knowledge can help as the registers are expected to be arranged in regular structures.
In the lack of such knowledge, boundaries can be determined by conducting image segmentation methods, e.g., the watershed transformation~\cite{meyer1994topographic}.
Besides, to reduce the impact of the noise, the two-dimensional (2-D) Wiener filter can be applied~\cite{lim1990two}, which can remove the noise by applying a pixel-wise adaptive low-pass Wiener filter to grayscale images.

After these steps, the attacker can choose two snapshots of cells as reference samples (i.e., templates): one containing \texttt{0}, and the other one \texttt{1} (such two different images can be easily found). 
Afterward, the attacker applies the cross-correlation over all the snapshots of register cells. 
Note that, since the positions of the individual register cells are given to the algorithm, the cross-correlation function (instead of the normalized one) can be employed to conduct the image registration. 
The reference sample that fits best to the targeted register cell determines the bit value contained in the snapshot.

\noindent
\textbf{\textit{Remark~1.}}
If giving labels to register cells on a training device is not feasible, the adversary still ends up with two groups of cells labeled as (\texttt{0},~\texttt{1}) or (\texttt{1},~\texttt{0}).
Consequently, she obtains two candidates for the secret key (verified by a single plaintext-ciphertext pair).
Therefore, having access to a training device is not an essential fact.

\noindent
\textbf{\textit{Remark~2.}}
The adversary should not necessarily look for the key registers. 
Recovering the state of the cipher -- either at initial cipher rounds when the input is known or at final cipher rounds when the output is known -- would suffice to reveal the key completely or partially, depending on the underlying cipher.
For example, having the state of AES-128 encryption at the first round (after AddRoundKey, SubBytes, or MixColumns) is enough to recover the entire 128-bit key, but for AES-256, two consecutive rounds should be covered.

\subsection{Scenario~2: Unknown Register Locations}
\label{sec:approach_localization}

If the location of the registers in the underlying design is not known to the adversary, the attack seems to be nontrivial.
Our proposal in such a case is to follow a two-step approach: i) distinguishing the registers from combinatorial cells, and ii) making use of a SAT solver to reveal the location of registers of interest, and finally, extracting the secret. 

\subsubsection{Identifying register cells}\label{sec:approach_localization:indentify}
To localize all register cells of the design on a chip, we propose an approach that takes advantage of the difference between sequential and combinatorial logic.
In synchronous designs -- as the most common design architecture -- every register is driven by the system clock\footnote{In case of clock gating, it should be made sure that the clock is propagated at the target cycle. A detailed discussion is given in Section~\ref{sec:discussion:clockgating}.}.
Consequently, all register cells have a clock input transistor.
In contrast, combinatorial logic is data driven, and thus has no clock input.
By conducting a traditional EOFM measurement at the clock frequency, the adversary can localize those clock input transistors.
The identified areas are the candidates for the location of register cells.
Furthermore, in those areas, conducting LLSI experiments with different data might give hints on the existence of a register.
In doing so, if the attacker finds at least one register cell, she can attempt to find similarities between its corresponding area and other candidate regions identified by an optical image or the LLSI image. 
Clearly after this step, the procedure of the automatic extraction of bit values from the snapshots, as explained in Section~\ref{sec:approach_direct}, can be followed.  

\subsubsection{Using SAT solver}\label{sec:scenario:sat}
Here, we suppose that the registers are distinguished from the other cells (e.g., through the technique given above), and their values can be recovered at multiple clock cycles, following the above instructions. 
We also suppose that the design architecture %
is known to the adversary, i.e., what is processed and stored at every clock cycle.
However, it is not known to the adversary which recovered value belongs to which register cell.
Having the above assumptions in mind, we propose to use a SAT solver to conduct the attack.
It is noteworthy that SAT solvers have also been used to construct algebraic side-channel attacks~\cite{renauld2009algebraic,oren2012algebraic}, where a SAT is written based on, e.g., the Hamming weight of the intermediate values recovered by a Template attack.
We made use of CryptoMiniSat~5~\cite{DBLP:conf/sat/SoosNC09}, which, compared to other alternatives, can more easily deal with XOR clauses. 

We first focus on a single snapshot at a certain clock cycle leading to binary observations denoted by $\{\omega_0,\ldots,\omega_{n-1}{\in\mathbb{F}_2}\}$ corresponding to $n$ registers of the design.
Some registers belong to the control logic (finite-state machine), which are out of our interest.
Therefore, we target $m \leq n$ registers according to the architecture of the underlying design.
For example, $m=256$ for an unprotected implementation of AES (128 bits for the state register and 128 bits for the key register).
If we define variables $v_{i\in\{0,\ldots,m-1\}}$ for the value of targeted register cells at the selected clock cycle, we can write
\vspace{-2mm}
\begin{equation}
v_i=c^i_0\omega_0 + ... + c^i_{n-1}\omega_{n-1},
\label{eq:var}
\end{equation}
where with $c^i_j$ we denote binary coefficients.
Since only one of the observations is associated to the $i$-th register cell, only one of the coefficients $c^i_{j\in\{0,\ldots,n-1\}}$ is 1, and the rest are 0.
In other words, $\forall i, \sum\limits_{\forall j}{c^i_j}=1$.
These are the first formulations that we require to include in the Boolean satisfiability problem~(SAT), which are generated individually for each targeted register cell $v_{i\in\{0,\ldots,m-1\}}$, and are independent of the observations $\omega$ and the architecture of the circuit under attack.

We should also add the formulations for~\eqref{eq:var} for each $v_i$.
Those observations $\omega_j$ that are 0 cancel out the corresponding coefficient $c_j$.
Therefore, we can write
\vspace{-1mm}
\begin{equation}\label{eq:varmain}
v_i \oplus \Bigl(\sumsc\limits_{\forall j, \omega_j=1}{c_j}\Bigr)=0.
\end{equation}
Having more snapshots at different clock cycles, the clauses for~\eqref{eq:varmain} should be repeated for $m$ distinct register variables $v_i$ based on the corresponding observations $\omega_j$.
However, the coefficients $c^i_j$ stay the same, i.e., they are defined only once for the entire circuit independent of the number of snapshots.

\begin{figure*}
	\centering
	\subfloat [DUT under the PHEMOS-1000 FA microscope]{
		\includegraphics[height=.3\textwidth]{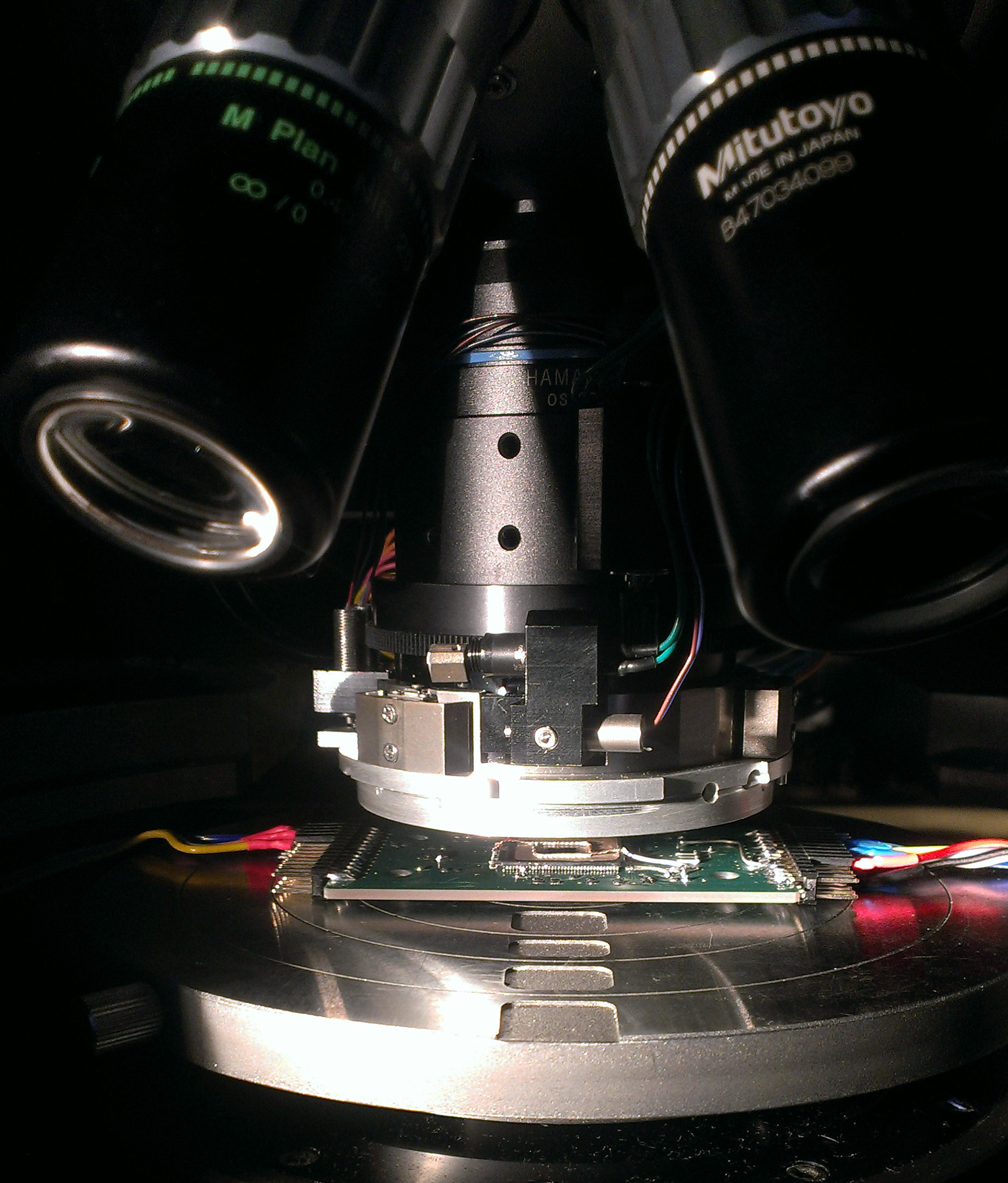}
		\label{fig:setup_DUT_photo}}
	\hfill
	\subfloat[Laser scan image of the DUT backside (5x lens)]{
		\scalebox{1.1}{
		\begin{tikzpicture}
			\node[inner sep = 0] (img) {\includegraphics[height=.273\textwidth]{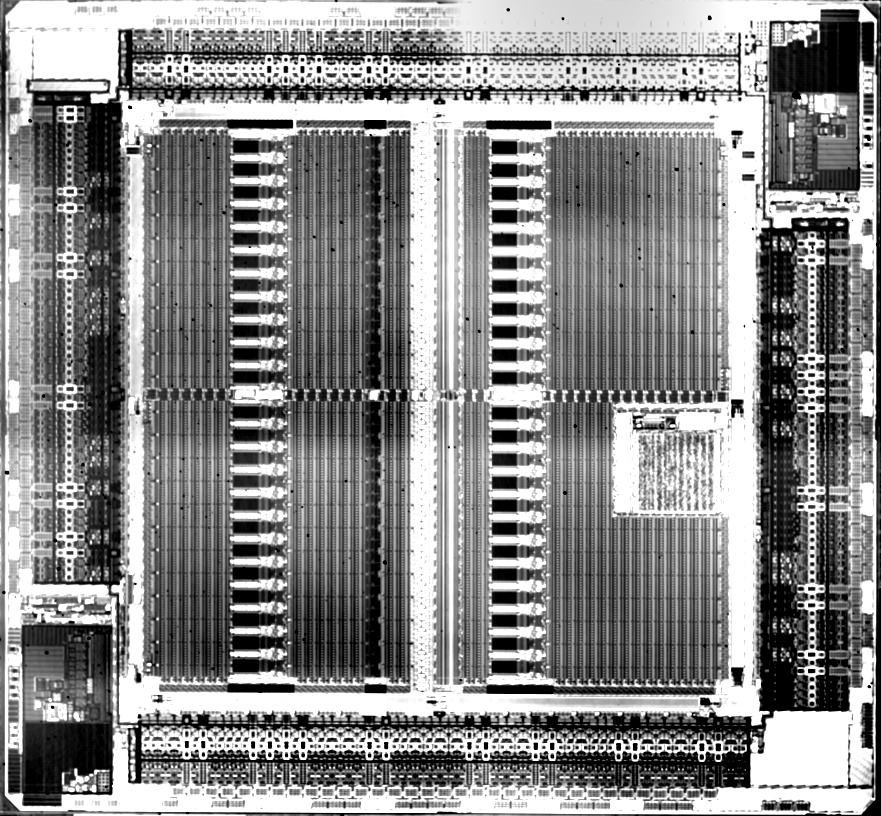}};
			\draw[red, very thick] ($(img.south west) + (3.66,0.9)$) rectangle ($(img.south west) + (3.92,1.25)$);
		\end{tikzpicture}
		}
		\label{fig:setup_DUT_laser}}
	\hfill
	\subfloat[Zoom-in of the framed area containing the LABs (50x lens)]{
		\includegraphics[height=.3\textwidth]{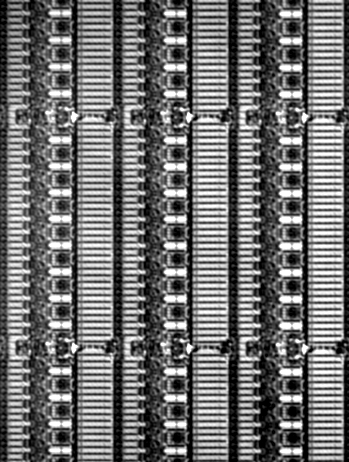}
		\label{fig:setup_DUT_laser_LABs}}
	\setlength{\abovecaptionskip}{2pt}
	\setlength{\belowcaptionskip}{5pt} %
	\caption{Device under test (DUT): Intel Cyclone IV FPGA with part number EP4CE6E22C8N.}
	\label{fig:setup_DUT}
	\vspace{-2mm}
\end{figure*}
The remaining task is to link the variables $v_i$ (of targeted register cells) at different clock cycles.
This is done based on the underlying design architecture of the circuit under attack and the functions it realizes.
For example, in a round-based architecture, the state register cells store the output of the cipher round function, and the key register cells the round keys.
In a serialized architecture, the content of the registers is shifted (e.g., in a byte-wise fashion), and certain operations (e.g., Sbox) are applied on particular registers at determined clock cycles.
We will elaborate an example in Section~\ref{sec:results_unknown}.

For a masked implementation with $d+1$ shares, the number of targeted registers at each clock cycle is $m(d+1)$ (e.g., $512 \times 2$ for a first-order masked implementation of AES using the state and key registers with 2 shares).
Therefore, the entire formulations given in~\eqref{eq:varmain} should be repeated $d+1$ times.
In the next step, we define $m$ virtual variables $\nu_i= \bigoplus\limits_{l=1}^{d+1}v_{i,l}$ (for each clock cycle), where $(v_{i,1}, \ldots, v_{i,d+1})$ represent variable $\nu_i$ with $d+1$ shares.
The corresponding formulations should be also added to the SAT.
The rest is similar to an unmasked implementation, i.e., the (unmasked) variables $\nu_i$ at different clock cycles are linked based on the design architecture of the circuit under attack.
We give a detailed explanation how to write the clauses in Appendix~\ref{appendix_clauses}.
\vspace{-0.5mm}
\section{Experimental Setup}
\label{sec:setup}
To evaluate our proposed attack, we need a target device that can run masked AES implementations of different protection orders.
In order to conduct LLSI, the power supply of the device must be modulated, and the backside of the chip must be optically accessible. 
Since snapshots of a large number of registers in multiple clock cycles have to be acquired, the automation of LLSI measurements would be beneficial.

\subsection{Device Under Test (DUT)}
\label{sec:setup_dut}
Our target device was an Intel Cyclone IV FPGA~\cite{alteracor_cyclone_2016} (see Fig.~\ref{fig:setup_DUT}). It is manufactured in a $60$\,\si{\nano\meter} technology and contains $392$ logic array blocks (LABs), each consisting of $16$ logic elements (LEs). The LEs mainly consist of a four-input look-up table (LUT) and a programmable register. Furthermore, in every LE, there is logic for loading and clearing data, routing, and clocking.
To access the backside of the chip, we opened the package and thinned the bulk silicon to a remaining depth of around $25$\,\si{\micro\meter}.%
We soldered the prepared sample upside down to a custom Printed Circuit Board (PCB) to expose connections to input/output and power supply pins.
To keep the power supply modulation for LLSI as unaffected as possible, we did not place capacitors on the PCB.

\subsection{Electrical and Optical Setup}
\label{sec:setup_electrical}
\begin{figure}
    \centering
	{\small
	\def\svgwidth{\linewidth}
    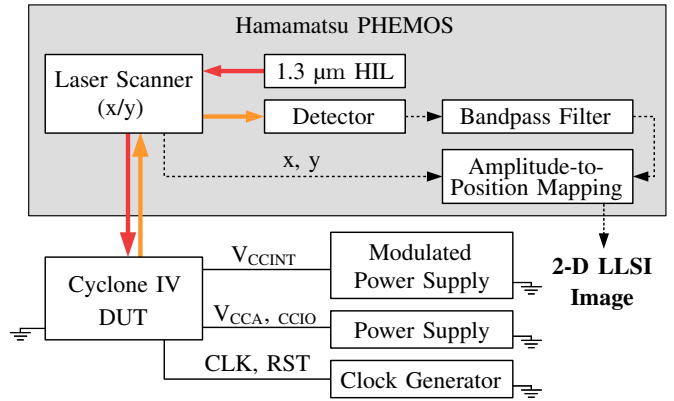}
	\setlength{\abovecaptionskip}{-12pt}
	\setlength{\belowcaptionskip}{-4pt} %
    \caption{Electrical and optical setup for conducting LLSI experiments.}
    \label{fig:setup_electo-optical}
	\vspace{-5mm}
\end{figure}

As the setup (Fig.~\ref{fig:setup_electo-optical}), we used a Hamamatsu PHEMOS-1000 FA microscope with optical probing capabilities.
It is equipped with a $1.3$\,\si{\micro\meter} high-power incoherent light source (HIL) and 5x/$0.14$NA, 20x/$0.4$NA, and 50x/$0.71$NA objectives.
An additional scanner-zoom of 2x, 4x and 8x is available.
For EOFM/LLSI, the laser is scanned over the device using galvanometric mirrors, and the reflected light is separated by semi-transparent mirrors and fed into a detector.
Its output is then fed into a bandpass filter set to the frequency of interest.
The resulting amplitude at every scanning location is mapped to its position and displayed as a grayscale encoded 2-D~image.

For LLSI, the supply voltage has to be modulated.
Therefore, the internal core voltage (V\textsubscript{CCINT}) of the DUT is supplied with $1.2$\,\si{\volt} by a Texas Instruments voltage regulator (TPS7A7001), whose feedback path is used to modulate the output voltage.
The sine wave signal used for this purpose is generated by a Keithley 3390 function generator, and a Toellner laboratory power supply (TOE8732) provides the DC voltage.
An LLSI peak-to-peak modulation amplitude up to $700$\,\si{\milli\volt\peakpeak} at $90$\,\si{\kilo\hertz} is possible without disturbing the functionality of the device.
The auxiliary voltage pin (V\textsubscript{CCA}) and I/O voltage pin (V\textsubscript{CCIO}) are supplied by the second channel of the TOE8732, which is set to $2.5$\,\si{\volt}.
The clock for the DUT is externally supplied via a Rigol DG4162 function generator, which allows single-stepping and stopping the clock.

\vspace{-1mm}
\subsection{Automation of LLSI Acquisition}
\label{sec:setup_llsiAutomation}
To create snapshots of the registers in multiple clock cycles in an automated fashion, we use the CadNavi interface provided by the PHEMOS-1000 and the USB interface of the clock generator.
The CadNavi interface gives access to functionalities of the PHEMOS, e.g., moving the microscope stage, adjusting the focus, and starting and stopping the measurements.
Using the clock generator, the DUT can be reset, and clock cycles can be advanced in single steps.
In the LabView programming environment, we implemented a scanning routine as follows.
First, the device is stopped at the clock cycle of interest.
The stage is then moved to a location of interest, where the focus is adjusted, and drift of the optical setup in \textit{x}- and \textit{y}-direction is corrected.
For drift correction, we apply an elastic image registration on the current optical image and an image recorded before the first measurement.
Finally, an optical image is taken and the LLSI snapshot is gathered.
After the program has gathered snapshots of all locations of interest, the same procedure begins for the next clock cycle.

\section{Results}
\label{sec:results}

\subsection{Data Dependency of LLSI Measurements}
\label{sec:results_singleBit}
\begin{figure}
	\centering
	\scalebox{1.15}{
		\import{figs/results/eofm_localization/}{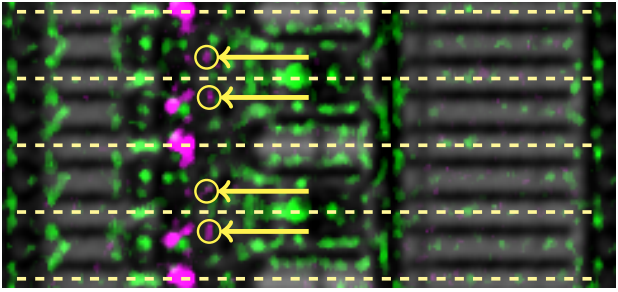}
	}
	\caption{Experiment for identifying the register cells. EOFM image at the clock frequency (magenta) and LLSI signature (green), overlaid onto an optical image and gathered in parallel while the device was running. LE boundaries indicated by dashed lines and potential clock transistors of registers by arrows. \vspace{-15pt}
	}
	\label{fig:results_localizationEOFM}
\end{figure}

To find the approximate register locations on the FPGA, we first conducted an EOFM measurement at the clock frequency~\cite{rahman2019key}, while the device was operating normally.
In the result shown in Fig.~\ref{fig:results_localizationEOFM}, we could identify several spots switching at the clock frequency.
We presume that some of the spots are the actual clock buffers for the registers, and others are part of the clock routing buffers between the LEs. 
By comparing the chip layout from the FPGA design software with the optical image, we identified the horizontal boundaries between the LEs, as indicated with the dashed lines in Fig.~\ref{fig:results_localizationEOFM}.
Note that every second LE seems to be flipped horizontally.
We then identified clock activity spots that are at the same relative position for every LE, see marked spots in Fig.~\ref{fig:results_localizationEOFM}.
Because every LE contains only a single-bit register, we expected the registers to reside in the vicinity of these spots.

\begin{figure*}
	\centering
	\import{figs/results/single-bit-diff-llsi/}{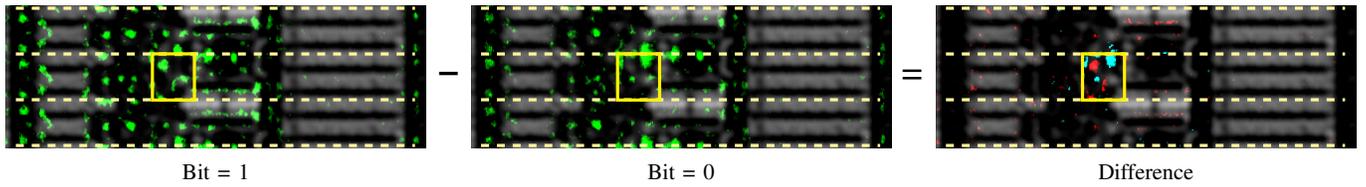}
	\setlength{\abovecaptionskip}{-15pt}
	\caption{LLSI measurement of 3 LEs (separated by dashed lines) with only the register of the centered LE (yellow window) set first to \texttt{1} and then to \texttt{0}, while keeping the other registers set to \texttt{0}. When subtracting images from each other, the result indicates the differences. Only the register at the centered LE shows a clear difference, indicating that the bit value has changed.\vspace{-10pt}}
	\label{fig:results_single_llsi}
\end{figure*}

To find a data dependency in the LLSI measurements and confirm the register location hypothesis, we targeted a single register cell.
For this, we set all surrounding registers to \texttt{0} and took two LLSI snapshots, one with the targeted bit set to \texttt{1}, and one with \texttt{0}, see Fig.~\ref{fig:results_single_llsi}. 
We set the modulation of V\textsubscript{CCINT} to $530$\,\si{\milli\volt\peakpeak} at $90$\,\si{\kilo\hertz} and scanned using the 50x lens with 2x zoom and a pixel dwell time of $10$\,\si{\milli\second/\pixel}.
Note that we could see a signature on the LLSI measurements already with a lower modulation amplitude, but we chose these settings to increase the SNR, and hence, decrease the scanning time.


By subtracting the captured LLSI measurements, the areas with differences become visible.
It can be observed that there is only one LE with differences, indicated by the yellow window in Fig.~\ref{fig:results_single_llsi}.
The size of this area is about $7$\,\si{\micro\meter}\,$\times$\,$9$\,\si{\micro\meter}, and located directly to the right of the potential clock buffer.
Due to the number of different spots, we assume that the window contains more than just the register.
Presumably, the in- and output transistors, as well as other logic, also contribute to the LLSI signature; however, this is irrelevant to our attack as its goal is to extract the bit values stored in the register cells.
To demonstrate how arbitrary data from the LLSI images can be read out, we took a snapshot of 24 registers containing randomly chosen data.
For an easier manual extraction, we have subtracted a reference snapshot with all registers set to \texttt{0}, see Fig.~\ref{fig:results_2LABs_diff} in Appendix~\ref{appendix_figures}.
Consequently, if there is a clear difference for a cell, it contains the value \texttt{1}; otherwise~\texttt{0}.

This leads us to the conclusion that the register inside the LAB and LE can be localized, and also the bit values \texttt{0} and \texttt{1} can be distinguished using a single LLSI measurement. 

\subsection{Implementation Under Attack}
\label{sec:setup_implementation}

We chose the AES-DOM implementation~\cite{gross_domainoriented_2016}, which is available on GitHub~\cite{h.gross_dom_2016}.
It is a serialized AES encryption engine that is given the shares of 128-bit plaintext and key, shifted in byte-by-byte during the first 16 clock cycles.
The code is written so that it allows the user to arbitrarily adjust the protection order (i.e., the number of shares), meaning that for a $d+1$ sharing scheme, it is expected to provide security against attacks up to $d$-th order by means of $d+1$ shares.
It requires a high number of random masks refreshed at every clock cycle, i.e., $(d+1)(9d+10)$ bits for $d+1$ shares.
Due to its serialized architecture, only one instance of the (masked) Sbox is instantiated.
Since the Sbox has 4 stages of pipeline intermediate registers (essential for almost any hardware masked implementation), a complete SubBytes operation takes 16+4 clock cycles.
MixColumns is also performed column-wise, requiring 4 clock cycles.
However, due to an interleaved fashion (ShiftRows and MixColumns being applied in parallel to SubBytes), the entire encryption can be terminated after 200 clock cycles~\cite{gross_domainoriented_2016}.

For the implementation on the FPGA, we restricted the AES-DOM core to be placed in a dedicated area on the FPGA using the logic fencing feature of the FPGA design software.
Our wrapper module, which is responsible for providing all inputs to the AES core, can thus be excluded from the hardware snapshots.
The highest protection, which we could fit on the FPGA (with our co-existing wrapper module), was of $4$\textsuperscript{th} order, resulting in 5 shares.

\subsection{Key Extraction with Known Register Locations}
\label{sec:results_known}

In the first scenario, we target a $d+1=3$-share\footnote{In the AES-DOM code~\cite{h.gross_dom_2016}, the protection order $d$ is shown by parameter $N=d$.} and a $d+1=5$-share implementation of AES-DOM (as given in Section~\ref{sec:setup_implementation}), resulting in $3 \times 128 = 384$ and $5 \times 128 = 640$ bits of key registers, respectively.
We placed all key registers to known locations.
To minimize the LLSI scanning time, we considered $3$ and $5$-share implementations occupying in total 24 and 40 LABs (each LAB with 16 register cells), respectively.  
As the input key shares are provided byte-by-byte to the AES-DOM core, after 16 clock cycles all key shares are stored inside the key registers; hence it is sufficient to extract the key register content only in the 16\textsuperscript{th} clock cycle.

We could achieve a reasonable SNR for the LLSI measurements with the 50x lens, 2x zoom, and a pixel dwell time of $3.3$\,\si{\milli\second/\pixel} with a V\textsubscript{CCINT} modulation of $640$\,\si{\milli\volt\peakpeak} at $90$\,\si{\kilo\hertz}.
Our scanning routine -- including autofocus and drift correction -- needs $2.7$ minutes to scan one LAB (containing 16 register cells).
Note that we scanned only the part of the LABs holding the register cells.
Scanning all $3$ and $5$-share key registers took around $65$ and $108$ minutes, respectively.

\begin{figure}
    \centering
	\includegraphics[width=.95\linewidth]{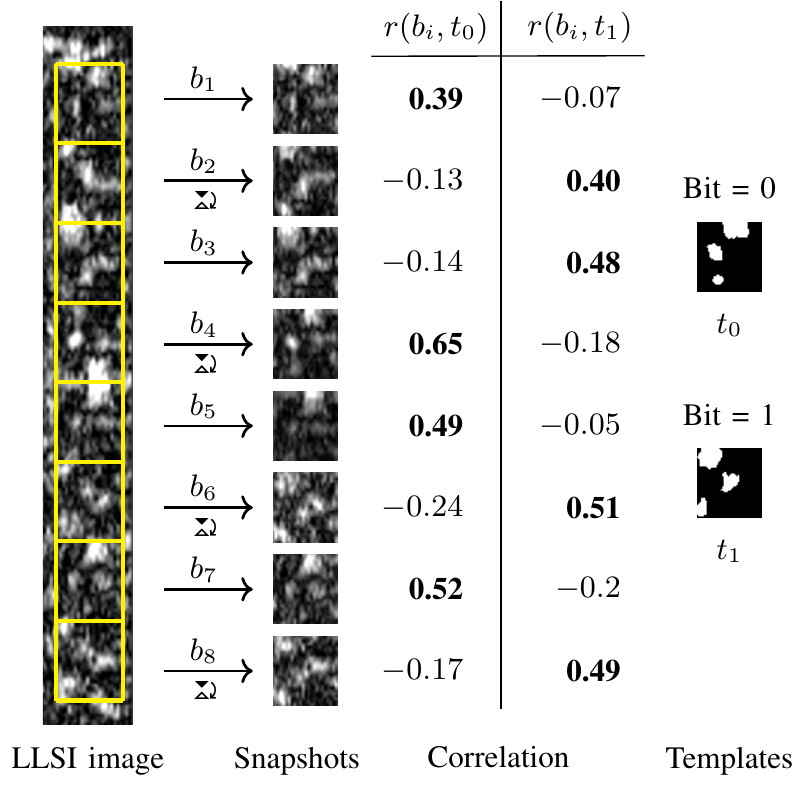}
	\setlength{\abovecaptionskip}{-5pt}
    \caption{Correlation-based data extraction mechanism from snapshots of half a LAB (8 bits). Due to the FPGA layout, every second cell has to be flipped horizontally. The correlation coefficient $r(a,b)$ between each cell and the templates for value \texttt{0} and \texttt{1} is calculated. The extracted bit value is determined by the template matching best.\vspace{-10pt}
	}
    \label{fig:results_knownRegs}
\end{figure}

We could easily read out the bit values from the LLSI measurements (even manually possible, for example, see Fig.~\ref{fig:results_knownRegs}).
Subtracting a reference measurement when zero stored in the registers (recorded, e.g., directly after resetting the device) could potentially facilitate manual readout, as also already observed in Section~\ref{sec:results_singleBit}.
However, we used an automated correlation-based extraction scheme which does not require to take snapshots of all registers while they contain zeroes.

\noindent
\textbf{\textit{Extracting bit values from snapshots.}}
To extract the bit values from the LLSI images as described in Section~\ref{sec:approach}, we applied off-the-shelf image processing algorithms provided in the Matlab software package~\cite{Matlabweb}.
First, we registered all the optical images that had been captured along with the snapshots using an elastic transformation. 
Note that here the process of registration refers to the transformation of the sets of data into one coordinate system, which should not be confused with the technique that we apply to identify the register values. 
The alignment enables us to cut every register cell according to the boundaries observed in Section~\ref{sec:results_singleBit} from the snapshot images in an automated fashion.
From the resulting cells, we chose two template snapshots of a single register cell for different bit values and subtracted them from each other to remove the signatures not representing the bit value. 
Then, as explained in Section~\ref{sec:approach_direct}, we applied noise reduction through adaptive filtering, and finally converted the templates to a binary mask, see Fig.~\ref{fig:results_knownRegs}. 
To extract the bit values, we calculated the \mbox{2-D} cross-correlation between the snapshot and each template. 
For determining the value of the register cells, the template for which the maximum correlation is achieved is taken into account. 
In our experiment, we extracted the value of all registers from the snapshots with 100\% accuracy. 
It is worth mentioning that for our approach, solely a pair of reference cells is required, which can be prepared straightforwardly. 
The efficiency of our technique should be evident when comparing it with machine learning methods that require a relatively large set of labeled cells. 

\begin{figure}
	\centering
	\includegraphics[width=.94\linewidth]{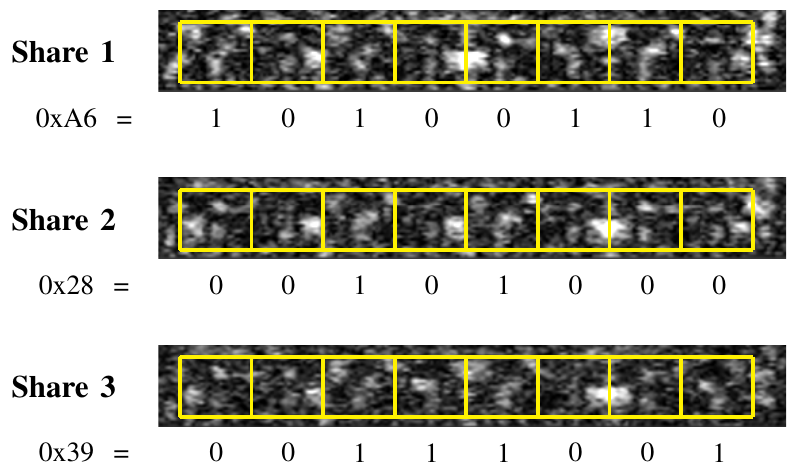}
	\setlength{\abovecaptionskip}{-2pt}
	\caption{
		Extracted values of the first byte of key register shares fo the $3$-share implementation. XOR'ing the results \texttt{0xA6}\,$\oplus$\,\texttt{0x28}\,$\oplus$\,\texttt{0x39}\,$=$\,\texttt{0xB7} reveals the first byte of the unshared key beginning with \texttt{0xB7FCBFF83...}
	\vspace{-15pt}}
	\label{fig:results_knownRegs_firstByte}
\end{figure}

Due to the underlying 2\textsuperscript{nd}- and 4\textsuperscript{th}-order Boolean masking scheme, by bit-wise XOR'ing all shares, the entire 128 bits of the AES key are trivially revealed (for the first key byte of the $3$-share implementation, see Fig~\ref{fig:results_knownRegs_firstByte}).
The raw LLSI measurements and extraction scripts for all experiments are available online as open-access research data\footnote{
\url{http://dx.doi.org/10.14279/depositonce-10440}}.

\subsection{Key Extraction with Unknown Register Locations}
\label{sec:results_unknown}

In the second scenario, we selected a $d+1=2$-share implementation of AES-DOM as the target.
We adjusted the size of logic blocks so that nearly all 16 registers in each LAB are used, occupying in total 45 LABs.
Note that these LABs cover the entire registers of the AES-DOM design, including the shared key registers, shared state registers, the intermediate masked Sbox registers, and those of finite-state machines.
However, we do not have any knowledge about the exact location of each register cell and enforce no other placement rule rather than what is explained above.
Using the scheme explained in Section~\ref{sec:approach_localization:indentify}, we localized the physical area on the chip where the register cells are placed, see Fig.~\ref{fig:results_localizationEOFM}. %

\newlength{\parsave}
\setlength{\parsave}{\parindent}
\setlength{\parindent}{0cm}
\begin{table*}
\caption{State of the registers of the AES-DOM design in the first 36 clock cycles, each row represents a register byte,\newline K: key bytes, S: SubBytes output, M: MixColumns output, K': 2nd round key bytes, S': 2nd-round SubByte output.}
\label{tab:arch_AES_DOM}
\begin{minipage}{0.348\textwidth}
\begin{adjustbox}{width=\textwidth,height=5cm,keepaspectratio}
\begin{lstlisting}[basicstyle=\linespread{0.9}\tiny]
1  2  3  4  5  6  7  8  9  10 11  12  13  14  15 
-  -  -  -  -  -  -  -  -  -  -   -   -   -   -  
-  -  -  -  -  -  -  -  -  -  -   -   -   -   K0 
-  -  -  -  -  -  -  -  -  -  -   -   -   K0  K1 
-  -  -  -  -  -  -  -  -  -  -   -   K0  K1  K2 
-  -  -  -  -  -  -  -  -  -  -   K0  K1  K2  K3 
-  -  -  -  -  -  -  -  -  -  K0  K1  K2  K3  S0 
-  -  -  -  -  -  -  -  -  K0 K1  K2  K3  S0  S1 
-  -  -  -  -  -  -  -  K0 K1 K2  K3  S0  S1  S2 
-  -  -  -  -  -  -  K0 K1 K2 K3  S0  S1  S2  S3 
-  -  -  -  -  -  K0 K1 K2 K3 S0  S1  S2  S3  S4 
-  -  -  -  -  K0 K1 K2 K3 S0 S1  S2  S3  S4  S5 
-  -  -  -  K0 K1 K2 K3 S0 S1 S2  S3  S4  S5  S6 
-  -  -  K0 K1 K2 K3 S0 S1 S2 S3  S4  S5  S6  S7 
-  -  K0 K1 K2 K3 S0 S1 S2 S3 S4  S5  S6  S7  S8 
-  K0 K1 K2 K3 S0 S1 S2 S3 S4 S5  S6  S7  S8  S9 
K0 K1 K2 K3 S0 S1 S2 S3 S4 S5 S6  S7  S8  S9  S10

-  -  -  -  -  -  -  -  -  -  -   -   -   -   -  
-  -  -  -  -  -  -  -  -  -  -   -   -   -   K0 
-  -  -  -  -  -  -  -  -  -  -   -   -   K0  K1 
-  -  -  -  -  -  -  -  -  -  -   -   K0  K1  K2 
-  -  -  -  -  -  -  -  -  -  -   K4  K5  K6  K7 
-  -  -  -  -  -  -  -  -  -  K4  K5  K6  K7  K4 
-  -  -  -  -  -  -  -  -  K4 K5  K6  K7  K4  K5 
-  -  -  -  -  -  -  -  K4 K5 K6  K7  K4  K5  K6 
-  -  -  -  -  -  -  K4 K5 K6 K7  K8  K9  K10 K11
-  -  -  -  -  -  K4 K5 K6 K7 K8  K9  K10 K11 K8 
-  -  -  -  -  K4 K5 K6 K7 K8 K9  K10 K11 K8  K9 
-  -  -  -  K4 K5 K6 K7 K8 K9 K10 K11 K8  K9  K10
-  -  -  -  -  -  -  -  -  -  -   -   -   -   -  
-  -  -  -  -  -  -  -  -  -  -   -   -   -   K12
-  -  -  -  -  -  -  -  -  -  -   -   -   K12 K13
-  -  -  -  -  -  -  -  -  -  -   -   K12 K13 K14
\end{lstlisting}
\end{adjustbox}
\end{minipage}~
\begin{fminipage}{0.334\textwidth}
\begin{adjustbox}{width=\textwidth,height=5cm,keepaspectratio}
\begin{lstlisting}[basicstyle=\linespread{0.9}\tiny]
16  17  18  19  20  21  22  23  24  25  26  27 
K0  K1  K2  K3  S0  M1  M2  M3  S4  M5  M6  M7 
K1  K2  K3  S0  S5  M2  M3  S4  S9  M6  M7  S8 
K2  K3  S0  S1  S10 M3  S4  S9  S14 M7  S8  S13
K3  S0  S1  S2  S15 S4  S9  S14 S3  S8  S13 S2 
S0  S1  S2  S3  S4  S9  S14 S3  S8  S13 S2  S7 
S1  S2  S3  S4  S9  S14 S3  S8  S13 S2  S7  S12
S2  S3  S4  S5  S14 S3  S8  S13 S2  S7  S12 S1 
S3  S4  S5  S6  S3  S8  S13 S2  S7  S12 S1  S6 
S4  S5  S6  S7  S8  S13 S2  S7  S12 S1  S6  S11
S5  S6  S7  S8  S13 S2  S7  S12 S1  S6  S11 K'0
S6  S7  S8  S9  S2  S7  S12 S1  S6  S11 K'0 K'1
S7  S8  S9  S10 S7  S12 S1  S6  S11 K'0 K'1 K'2
S8  S9  S10 S11 S12 S1  S6  S11 K'0 K'1 K'2 K'3
S9  S10 S11 S12 S1  S6  S11 K'0 K'1 K'2 K'3 S'0
S10 S11 S12 S13 S6  S11 K'0 K'1 K'2 K'3 S'0 S'1
S11 S12 S13 S14 S11 K'0 K'1 K'2 K'3 S'0 S'1 S'2

K0  K1  K2  K3  K0  K1  K2  K3  K4  K5  K6  K7 
K1  K2  K3  K0  K1  K2  K3  K4  K5  K6  K7  K8 
K2  K3  K0  K1  K2  K3  K4  K5  K6  K7  K8  K9 
K3  K0  K1  K2  K3  K4  K5  K6  K7  K8  K9  K10
K4  K5  K6  K7  K4  K5  K6  K7  K8  K9  K10 K11
K5  K6  K7  K4  K5  K6  K7  K8  K9  K10 K11 K12
K6  K7  K4  K5  K6  K7  K8  K9  K10 K11 K12 K13
K7  K4  K5  K6  K7  K8  K9  K10 K11 K12 K13 K14
K8  K9  K10 K11 K8  K9  K10 K11 K12 K13 K14 K15
K9  K10 K11 K8  K9  K10 K11 K12 K13 K14 K15 K'4
K10 K11 K8  K9  K10 K11 K12 K13 K14 K15 K'4 K'5
K11 K8  K9  K10 K11 K12 K13 K14 K15 K'4 K'5 K'6
K12 K13 K14 K15 K12 K13 K14 K15 K'0 K'1 K'2 K'3
K13 K14 K15 K12 K13 K14 K15 K'0 K'1 K'2 K'3 K'4
K14 K15 K12 K13 K14 K15 K'0 K'1 K'2 K'3 K'4 K'5
K15 K12 K13 K14 K15 K'0 K'1 K'2 K'3 K'4 K'5 K'6
\end{lstlisting}
\end{adjustbox}
\end{fminipage}
\begin{minipage}{0.292\textwidth}
\begin{adjustbox}{width=\textwidth,height=5cm,keepaspectratio}
\begin{lstlisting}[basicstyle=\linespread{0.9}\tiny]
28  29  30  31   32   33   34   35   36
S8  M9  M10 M11  S12  M13  M14  M15  K'0
S13 M10 M11 S12  S1   M14  M15  K'0  K'1
S2  M11 S12 S1   S6   M15  K'0  K'1  K'2
S7  S12 S1  S6   S11  K'0  K'1  K'2  K'3
S12 S1  S6  S11  K'0  K'1  K'2  K'3  S'0
S1  S6  S11 K'0  K'1  K'2  K'3  S'0  S'1
S6  S11 K'0 K'1  K'2  K'3  S'0  S'1  S'2
S11 K'0 K'1 K'2  K'3  S'0  S'1  S'2  S'3
K'0 K'1 K'2 K'3  S'0  S'1  S'2  S'3  S'4
K'1 K'2 K'3 S'0  S'1  S'2  S'3  S'4  S'5
K'2 K'3 S'0 S'1  S'2  S'3  S'4  S'5  S'6
K'3 S'0 S'1 S'2  S'3  S'4  S'5  S'6  S'7
S'0 S'1 S'2 S'3  S'4  S'5  S'6  S'7  S'8
S'1 S'2 S'3 S'4  S'5  S'6  S'7  S'8  S'9
S'2 S'3 S'4 S'5  S'6  S'7  S'8  S'9  S'10
S'3 S'4 S'5 S'6  S'7  S'8  S'9  S'10 S'11

K8  K9  K10 K11  K12  K13  K14  K15  K'0
K9  K10 K11 K12  K13  K14  K15  K'0  K'1
K10 K11 K12 K13  K14  K15  K'0  K'1  K'2
K11 K12 K13 K14  K15  K'0  K'1  K'2  K'3
K12 K13 K14 K15  K'4  K'5  K'6  K'7  K'4
K13 K14 K15 K'4  K'5  K'6  K'7  K'4  K'5
K14 K15 K'4 K'5  K'6  K'7  K'4  K'5  K'6
K15 K'4 K'5 K'6  K'7  K'4  K'5  K'6  K'7
K'4 K'5 K'6 K'7  K'8  K'9  K'10 K'11 K'8
K'5 K'6 K'7 K'8  K'9  K'10 K'11 K'8  K'9
K'6 K'7 K'8 K'9  K'10 K'11 K'8  K'9  K'10
K'7 K'8 K'9 K'10 K'11 K'8  K'9  K'10 K'11
K'4 K'5 K'6 K'7  K'8  K'9  K'10 K'11 K'12
K'5 K'6 K'7 K'8  K'9  K'10 K'11 K'12 K'13
K'6 K'7 K'8 K'9  K'10 K'11 K'12 K'13 K'14
K'7 K'8 K'9 K'10 K'11 K'12 K'13 K'14 K'15
\end{lstlisting}
\end{adjustbox}
\end{minipage}
\end{table*}
\setlength{\parindent}{\parsave}

To conduct the attack, we first investigated the design architecture of the AES-DOM, being serialized with the state and key registers shifted byte-wise, as stated before. 
Table~\ref{tab:arch_AES_DOM} represents the content of 32 registers (consisting of 8 bits each) stored in consecutive clock cycles for the first 36 clock cycles, whereas the order of rows in the table is not of our interest. 
For example, the first row shows that the register that stored K0 at clock cycle 16, will hold K1, K2, K3, S0, M1, M2,~... in the next clock cycles.
We would like to highlight that it is a symbolic representation and independent of the masking order, e.g., K0 represents all $d+1$ shares of the first byte of the key.

It can be seen that in clock cycle 16, all registers are filled; a part of the state registers with SubBytes' output and the first quarter with 4 bytes of the key.
The key register is also fully filled by the given key, which precisely justifies why we targeted this clock cycle for the attack in the first scenario, see Section~\ref{sec:results_known}.
Here, we also started at clock cycle 16 and collected LLSI measurements of the entire 45 LABs in 12 consecutive clock cycles.
Each full snapshot in a clock cycle took around 2\,hours.
Using the fully automated setup developed for this purpose, which applies drift correction mechanisms, we collected all snapshots in 24\,hours without any manual interaction.
Using the correlation-based extraction technique (see Section~\ref{sec:results_known}), we extracted the values stored in all registers during the 12 clock cycles.

\vspace{1mm}
\noindent
\textbf{\textit{Using SAT solver.}}
To extract the key, we made use of CryptoMiniSat~5~\cite{DBLP:conf/sat/SoosNC09} and followed the technique explained in Section~\ref{sec:scenario:sat}.
We developed a program in C++ which receives i) the architecture of the underlying design as in Table~\ref{tab:arch_AES_DOM}, ii) the masking order $d$, iii) the number of covered clock cycles $n$, and iv) the value of registers extracted by snapshots at $n$ clock cycles.
The program generates a Boolean satisfiability problem~(SAT) to be solved by the SAT solver.
For the above case (i.e., $d=1$ and $n=720$ register bits in 12 clock cycles), the SAT led to 3\,650\,048 clauses on 717\,728 variables.
The SAT solver required 1\,hour and 47\, minutes to solve the problem and successfully report the revealed key.
Note that the SAT solver does not find a unique solution, but all of them lead to the same revealed key.
This is due to the underlying masking scheme, i.e., when representing a variable $x$ by 2 shares, the SAT solver makes a distinction between $(x_1,x_2)$ and $(x_2,x_1)$, while both of them lead to the unique unmasked value $x$. 
This holds for all masked variables in the SAT.
If there are $\lambda$ of such mask variables, the SAT solver can find $\left(\left(d+1\right)!\right)^\lambda$ correct solutions.

\begin{table*}
	\caption{The required time for the SAT solver to report a solution, successfully recovering the key, for different masking order $d$ and various number of covered clock cycles by snapshots. \vspace{-10pt}}
	\label{tab:SAT:result}
	\centering
	\renewcommand*{\arraystretch}{1.2}
	\setlength\tabcolsep{4pt}
	\noindent\begin{tabular}{@{\extracolsep{\fill}} *{13}{c|}c}
		\textbf{Masking}   & \multicolumn{13}{c}{Number of covered clock cycles starting from 16}\\\cline{2-14}
		\textbf{order} $d$ & 9 & 10 & 11 & 12 & 13 & 14 & 15 & 16 & 17 & 18 & 19 & 20 & 21\\\hline
		0 & 1.5\,h  & 7\,m    & 2\,m    & ~~54\,s &  ~~46\,s &  ~21\,s &  ~19\,s & ~~~24\,s &  ~~~19\,s &  ~~~17\,s &  ~~19\,s &  ~~~15\,s & ~~~~~9\,s\\
		1 & -       & -       & -       & 1.78\,h &  ~~14\,m &  ~10\,m &  ~~8\,m & ~~~~7\,m &  ~~~~8\,m &  ~~~~6\,m &  ~~~6\,m &  ~~~~5\,m &  ~~~~6\,m \\
		2 & -       & -       & -       & -       & 1.76\,h  &  ~56\,m &  ~47\,m & ~~~38\,m &  ~~~39\,m &  ~~~30\,m &  ~~28\,m &  ~~~26\,m &  ~~~21\,m\\
		3 & -       & -       & -       & -       & -        &  5.4\,h &  4.5\,h & ~~2.5\,h &  ~2.83\,h &  ~2.15\,h &  1.93\,h &  ~~1.8\,h &  ~~1.2\,h\\
		4 & -       & -       & -       & -       & -        & -       &  9.5\,h & ~8.91\,h &  ~7.71\,h &  ~6.16\,h &  5.65\,h &  ~4.75\,h &  ~4.71\,h\\
		5 & -       & -       & -       & -       & -        & -       &  1.1\,d & 20.61\,h &  17.96\,h &  16.08\,h &  18.5\,h &  21.55\,h &  19.11\,h\\
		6 & -       & -       & -       & -       & -        & -       & -       & ~~1.8\,d &  ~~1.9\,d &  ~1.75\,d &  ~1.8\,d &  ~1.49\,d &  ~1.35\,d
	\end{tabular}
\end{table*} 

\vspace{1mm}
\noindent
\textbf{\textit{Extension.}}
To examine the efficiency of this approach for different numbers of shares $d+1$ and different numbers of covered clock cycles $\eta$, we have conducted several investigations. 
We simulated the AES-DOM for $d\in\{0,\ldots,6\}$ and extracted all register values at the first 36 clock cycles (see Table~\ref{tab:arch_AES_DOM}).
Note that we supplied the implementation with random masks (refreshed at every clock cycle), and did not consider the name/order of registers when extracting their values.
Starting from clock cycles 16, we ran the SAT solver on SATs covering $\eta\in\{9,\ldots,21\}$ clock cycles, i.e., from clock cycle 16 to clock cycle 24 up to 36.
We repeated this experiment with 10 sets of different plaintext/key (and random masks).
We found out that the SAT solver usually needs less time to find the solution when more clock cycles are covered by the SAT (expected, as it contains more information).
We further recognized that there is a minimum number of required covered clock cycles depending on the number of shares. 
The averaged results obtained using a machine with a 2.6\,GHz CPU and 256\,GB RAM are shown in Table~\ref{tab:SAT:result}.
Note that multithreading is not beneficial here, as CryptoMiniSat~5 looks for different solutions by each thread.
Besides, starting before the clock cycle 16 is not helpful since some registers do not contain meaningful data (see Table~\ref{tab:arch_AES_DOM}).

We have also investigated other design architectures.
In short, if the circuit does not allow the collection of enough snapshots per encryption/decryption (e.g., at most 10 in a round-based AES-128 encryption), snapshots for more inputs (plaintexts) can be collected.
Although it becomes out of the single-trace feature of our attack, it still allows recovering the secrets by a few snapshots (corresponding to different inputs).
As a general overview, a design which requires a higher number of clock cycles for each encryption/decryption would also exhibit more information in the snapshots.
We should stress that due to their high area overhead, usually just one instance of some basic blocks (like Sbox) is instantiated in masked implementations, leading to a high number of clock cycles per encryption/decryption.
This would potentially decrease the number of required snapshots in our attack.

\section{Discussion}
\label{sec:discussion}

\subsection{Attack Feasibility}
\label{sec:discussion_feasibility}

\subsubsection{Clock control}
\label{sec:discussion_clock}
For taking a snapshot of registers in a region of interest, the registers' contents should not be updated by the clock signal.
Therefore, the adversary either needs to halt the clock signal for every snapshot or find a time window, where the registers' contents remain constant for several clock cycles, sufficient for taking a snapshot.
Depending on the hardware designer, the state of the (masked) registers might not be cleared after the termination of the encryption/decryption.
The same observation has been reported in~\cite{moos_static_2019}.
In such cases, there is no need to have any control over the clock.
If the locations of the registers are known to the adversary, a snapshot from all key registers after the encryption/decryption can be taken to recover the key.
However, as multiple snapshots from successive clock cycles are required for the scenario with unknown register locations, this method cannot be applied.
Thus, controlling the clock signal is inevitable.
To stop the clock, we have identified the two following possible scenarios.

\noindent
\textbf{\textit{External clock.}}
In the most uncomplicated scenario, the clock is supplied to the chip externally.
Hence, the adversary can easily tamper with the clock signal before it enters the chip and keep it low/high at her desired periods to take a snapshot.
Naturally, she can repulse the clock again to move one or several clock cycles further with encryption/decryption.

\noindent
\textbf{\textit{Internal clock.}}
The attack becomes more challenging if the clock is generated internally on the chip.
Depending on the target platform (i.e., FPGA or ASIC), the attacker needs to apply more sophisticated techniques to tamper with the clock.
If the target is an SRAM-based FPGA, the attacker can use laser fault injection to manipulate the clock source configuration (e.g., based on ring-oscillators) or its routing configuration to stop the clock signalling~\cite{tajik2015laser,lohrke2016automated}.
To take a snapshot of registers, the adversary first needs to inject a fault into the clock circuitry at her desired cycle and then take a snapshot.
However, the challenge would be to reactivate the clock for the next snapshots.
Although rebooting the FPGA leads to the correct reconfiguration and reactivation of the clock circuitry, it will not be helpful for the next snapshots due to newly generated random masks.
Although successive immediate fault injections are feasible in principle, it might be impractical due to laser setup limitations.
Moreover, laser fault injection is not effective in case of an ASIC or a flash-based FPGA since only transient faults can be injected, which is usually not sufficient to halt the internal clock permanently.

A more realistic solution, applicable to all platforms, is circuit editing using Focused Ion Beam~(FIB).
Using FIB, the attacker can physically cut the metal lines responsible for clock signal delivery or damage the transistors of clock buffers to stop the clock.
After disconnecting the internal clock from the cipher, the attacker can provide her own controllable external clock signal by injecting pulses into clock lines using active nano-probe needles~\cite{kleindiek}.
Even though FIB circuit editing is an invasive technique, it is a practically feasible approach~\cite{helfmeier2013breaking}.
Thus, we believe that an internal clock cannot stop the attacker from mounting snapshot attacks, although it increases the difficulties.

\subsubsection{Clock Gating}\label{sec:discussion:clockgating}
In synchronous circuits, clock gating can be deployed to reduce dynamic power consumption by cutting the clock signal from flip-flops when they are not in use.
In this case, since the clock signal is not continuously delivered to a specific group of registers, a question rises about the feasibility of conducting EOFM on an unknown layout to localize the registers.
To ensure that all clock gated registers are receiving the clock signal during an EOFM measurement, the dwell time of the laser at each pixel has to be larger than full encryption/decryption time.
As a result, we can be confident that the gated registers have been activated temporarily and received the clock signal.
Note that while the clock signals for these gated registers might not be periodic anymore during the dwell time of the laser, they still contain the clock frequency component, however with a lower amplitude.
Therefore, an EOFM measurement with the clock frequency reveals clock buffers of gated registers with different modulation intensities, i.e., stronger modulation for always active registers and weaker modulation for gated registers.
For instance, assume that the cryptographic core is running with a 100\,\si{\mega\hertz} clock, and the dwell time of the laser is 1\,\si{\milli\second\per\pixel}.
In this example, AES DOM requires about 200 clock cycles or 2\,\si{\micro\second} to complete an encryption.
Hence, by keeping the cryptographic operation in a loop during an EOFM measurement, the AES circuit finishes the encryption 500 times while the laser beam is still at the same position.
Upon the laser's movement to the next pixel, the same number of operations in the loop occurs until the entire die is scanned with the laser.
Thus, by setting the correct relation between the clock frequency and the dwell time for the laser, all registers still can be localized while clock gating is in use.
Note that gates involved in the combinatorial logic will not be falsely identified as clock buffers, because they are updated only on either the rising or falling edge of the clock signal while the clock buffers toggle on both edges.
Therefore, the combinatorial gates -- except those belonging to the clock tree -- do not appear on the EOFM image.

\subsubsection{Time expenditure and Attack Cost}
\label{sec:discussion_cost}
One might argue that the time-consuming task of taking the snapshots discourages an adversary from mounting the attack, especially if all registers have to be covered in several clock cycles.
For the 2-share implementation, it took $24$\,hours to capture snapshots of all registers in 12 clock cycles, see Section~\ref{sec:results_unknown}.
The time fraction for a single LAB (16 registers) is $2.67$\,\si{\minute}.
Note that autofocus and drift correction significantly contribute to that time.
However, the LLSI scan, which creates the actual snapshot of the registers, takes only around $65$\,\si{\second}.
Therefore, using a more stable optical setup, the acquisition time could potentially be reduced by up to 60\%.
Furthermore, the registers on the used FPGA are spread over the device with much space in between.
On an ASIC implementation, the registers are potentially placed closer together, and thus, a smaller area needs to be imaged by LLSI.
Nevertheless, we consider the measurement time of our setup not as a hurdle for an attacker, because the measurements are fully automated and hence can run unsupervised without the presence of an operator.
Therefore, we think that -- concerning measurement time -- our approach is practically feasible in a real scenario.

While laser scanning microscopes are not as cheap as typical oscilloscopes for power/EM analysis, they are common FA equipment.
They can be rented for about $\$300$/h including an operator from different FA labs.
Therefore, depending on the attack scenario, one can estimate the cost of such attacks based on the number of shares and the size of the die.
For instance, the estimated cost to perform LLSI for the known layout of 3-share and 5-share masked AES implementations would be $\$325$ (65 min.) and $\$540$ (108 min.), respectively.
Naturally, the cost for an unknown layout would increase, since several snapshots from the entire die have to be taken.
However, the cost would increase only linearly by the number of registers on the chip.
The estimated cost to mount LLSI attack against an unknown layout with 2-share masked AES implementation would be $\$7200$ (24 hours).

\subsubsection{Optical resolution and register size}
\label{sec:discussion_resolution}
In the FA community, optical probing has been shown to be applicable even to the 10\,\si{\nano\meter} technology node by using a Solid Immersion Lens~(SIL), leading to an optical resolution of around 200\,\si{\nano\meter}~\cite{von2015optical, boit2016ic}. 
For smaller technology nodes, a higher resolution can be achieved in the visible light regime~\cite{beutler2015visible, boit2015contactless}.
For our experiments, we did not use an SIL; hence, the resolution is $\approx 1$\,\si{\micro\meter} due to the wavelength of the laser. 
This resolution might seem low for the DUT manufactured in a $60$\,\si{\nano\meter} technology.
However, unlike IC failure analysis, the security evaluation of ICs does not have to rely on targeting a single transistor; therefore, optical resolution requirements can be relaxed to a certain extent. 
The comparison of technology size and optical resolution often misleads to the assumption that optical probing is not possible for small technology sizes.
This has already been shown wrong in~\cite{tajik_power_2017}, where extracting the bitstream from a 28\,\si{\nano\meter} FPGA was demonstrated.

The size of the area which we used to extract the logic state of one register from, has a dimension of about $7$\,\si{\micro\meter}\,$\times$\,$9$\,\si{\micro\meter} for our DUT manufactured in a $60$\,\si{\nano\meter} technology. 
This area contains multiple transistors.
For traditional optical probing techniques, like EOP, the distance between transistors is critical for being able to extract the waveform from exactly one transistor and not a mixture of different signals.
However, for LLSI, it is not crucial whether the laser spot covers multiple transistors at a time or not.
As long as different signatures for different logic states can be observed in the LLSI measurements, the stored data can be extracted successfully.

\subsubsection{Chip preparation and silicon access}
\label{sec:discussion_preparation}
For our attack, we had to depackage the target chip and mount it upside-down on a customized PCB to establish access to the silicon backside.
This makes the attack semi-invasive, and one might argue that the effort for chip preparation puts a too high hurdle on the attacker.
However, note that modern chips are increasingly manufactured in flip-chip packages, due to performance, size, cost, and environmental compatibility reasons~\cite{tong_advanced_2013}. %
Here the silicon backside is directly exposed to the attacker, and no chip preparation is necessary.
Therefore, depending on the chip packaging, our attack can also be non-invasive, cf.~\cite{tajik_power_2017}.
\subsection{Theory vs. Practice}
It is tempting to claim that our results rule out the application of the $t$-probing model as presented in~\cite{ishai_private_2003}. 
In this regard, we highlight two main points. 
First, our attack falls only partially within that framework as it requires that the $t$ probes should not move within a time period. 
Second, but more interestingly, our results demonstrate that some of the assumptions made in~\cite{ishai_private_2003} do not always hold in reality. 
More concretely, in~\cite{ishai_private_2003} and its follow-up studies, the measure of the cost of a probing attack is associated with the value $t$, which is shown to be ineffective for our attack. 
For this purpose, for a practically feasible, yet more powerful adversary mounting our proposed attack, the spatial coverage and/or the resolution of the probe play a much more vital role. 
Moreover, it is claimed in~\cite{ishai_private_2003} that, even in the presence of a fully adaptive adversary moving the probes within a clock cycle, the security is guaranteed as long as the total number of probes in each clock cycle does not exceed $t$. 
Conversely, we present a powerful new attacker, who is not limited by the number of probes as long as she can manipulate the usual functionality of the clock, which is very likely as explained above and also practically demonstrated by us.
To sum this up, the existence of such powerful attackers suggests that the model presented in~\cite{ishai_private_2003} should be revisited.
Of course, the cost for such powerful attackers is higher than that for a classical SCA attack, and there is certainly a trade-off between the cost and the gain depending on the value of the secrets stored in the device.

\subsection{Potential Countermeasures}
\label{sec:discussion_countermeasures}
Our attack consists of four main steps, namely i) accessing the IC backside, ii) modulating the power supply, iii) scanning with a thermal laser, and iv) localizing the key/state registers.
Possible countermeasures can be designed and integrated into the chip to prevent each step.%
\subsubsection{Package-level countermeasures}
The optical access to the backside of the chip can be prevented after the fabrication and during the packaging of the die.
For instance, active backside coatings~\cite{amini2018assessment} can make the backside of the chip opaque to the laser scanning microscopy.
Since these coatings interact with the transistors, they can detect any tampering attempt.
Unfortunately, passive coating layers are not effective since they can be removed mechanically without any consequences.

\subsubsection{Device-level countermeasures}
To take a snapshot from the hardware, the core voltage of the device needs to be modulated with a specific frequency during the laser irradiation on the transistors.
For preventing the modulation of the supply voltage, internal voltage regulators can be integrated into the circuit to isolate the supply voltage of secure cores from the outside of the device and keep the core voltage constant.
Such regulators have already been proposed to defeat power and EM SCAs~\cite{kar2018reducing}.
As a side note, supplying a voltage regulator by a low voltage (close to its predefined output level) can lead to an unstable output or a transparency between input and output.
While the former case already might be sufficient for LLSI, in the latter case, the adversary becomes able to modulate the internal supply voltage at her will.
Moreover, distributed temperature sensors can be deployed on the die to detect local temperature variations resulting from the laser beam.
However, it should be noted that such temperature sensors have to operate independently from the main system clock; otherwise, they will also be deactivated by halting the clock.
Since the wavelength of the thermal lasers is larger than the bandgap of the silicon, no electron-hole pairs are generated upon the incident of photons, and therefore, conventional silicon-based light sensors do not trigger.
Temperature sensors can be either realized by timing-sensitive circuits (e.g., ring-oscillators~\cite{tajik2017pufmon}) or specific materials with longer bandgap wavelengths. 

\subsubsection{Circuit-level countermeasures}

A possible way to defeat our proposed attack is to change the location of registers dynamically.
It cannot be done physically, but it seems to be possible logically.
Suppose that every single bit is allowed to be stored in a set of $k$ registers.
Having $n$ bits, $k \times n$ register cells are required. 
In addition to this overhead, a mechanism is required to assign one of such $k$ register cells to a single-bit value, dynamically selected at every clock cycle, and independent of other single-bit values. 
Indeed, we need to randomize the location of registers, independent of any masking scheme integrated to defeat classical SCA attacks.  
Realizing this might be possible by a form of reconfigurability. 
To the best of our knowledge, there is no such a scheme known to the hardware security community, and therefore, it is among our planned future works.

\section{Conclusion}
\label{sec:conclusion}

Masking is the most effective protection for cryptographic implementations against (passive) SCA attacks.
The mathematical proof of the probing security models, however, assumes a limited number of probes available to the attacker.
This assumption holds for virtually all practically feasible SCA attacks reported so far.
We introduced a new optical attack approach that can capture \textit{hardware snapshots} of the IC's entire logic state.
It is a single-trace technique offering a number of probes that is only bounded by the number of transistors on the chip.
We showed that extracting the keys from 2-, 3- and 5-share AES-128 implementations is practically feasible, even when the exact register locations are not known to the attacker.
Due to the practically unlimited number of probes in our attack, implementations with higher protection orders (i.e., with a high number of shares) are vulnerable as well.
The complexity of the attack depends on the design architecture, the number of shares, and the knowledge of the adversary about the underlying implementation.
The results confirm (again) that cryptography should not rely on complexity of physical attacks.
Moreover, assumptions made in theoretical models can be invalidated through more advanced FA techniques, and hence, one should not underestimate them.
We believe that the integration of countermeasures to defeat our attack is not a trivial task.
Nevertheless, we gave an overview of the potential countermeasures at different levels of abstraction.

\section*{Acknowledgment}
The work described in this paper has been supported in part by the Einstein Foundation in form of an Einstein professorship - EP-2018-480, and in part by the Deutsche Forschungsgemeinschaft (DFG, German Research Foundation) under Germany's Excellence Strategy - EXC 2092 CASA - 390781972.
The authors would also like to acknowledge Hamamatsu Photonics K.K. Japan and Germany for their help and support on the PHEMOS system.
The authors declare no other financial and non-financial competing interests.

\ifCLASSOPTIONcaptionsoff
  \newpage
\fi

\bibliographystyle{IEEEtran}
\bibliography{bibtex/bib/IEEEabrv,bibtex/bib/IEEEexample,bibtex/bib/hw-snapshot}

\appendices
\pagebreak
\section{Additional Figure}
\label{appendix_figures}
\FloatBarrier
\begin{figure}[h!]
	\centering
	\import{figs/results/multiple-bits-diff/}{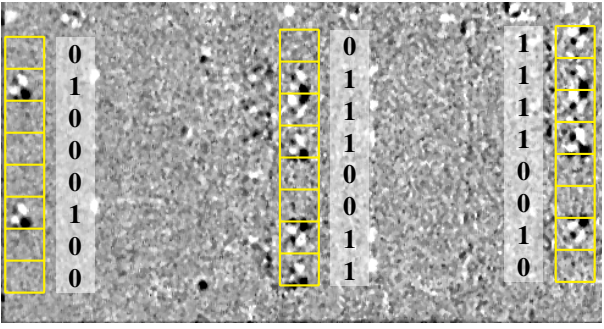}
	\caption{Difference image of a snapshot covering three times eight registers, once filled with random data, and once with zeroes. A significant difference (black and white spots) for a register corresponds to bit value \texttt{1}.}
	\label{fig:results_2LABs_diff}
\end{figure}
\FloatBarrier

\section{SAT Clauses}
\label{appendix_clauses}

We suppose that the registers are distinguished from the other cells (e.g., through the technique given in Section~\ref{sec:scenario}), and their values can be recovered at multiple clock cycles, following the given instructions. 
We also suppose that the design architecture %
is known to the adversary, i.e., what is processed and stored at every clock cycle.
However, the relation between the recovered values (through snapshots) and the register cells is unknown. 
In other words, it is not known to the adversary which recovered value belongs to which register cell. 

Having the above assumptions in mind, we use CryptoMiniSat~5~\cite{DBLP:conf/sat/SoosNC09} to conduct the attack, which, compared to other similar SAT solvers, can more easily deal with XOR clauses. 
We should highlight that in such SAT solvers, the problem should be written in Conjunctive Normal Form~(CNF), or let say product of sums.
Each clause is a sum (logical OR) of a couple of variables (or their invert).
The product (logical AND) of all clauses should be \texttt{True}, hence every clause should be \texttt{True}.
CryptoMiniSat allows us to easily define XOR-based clauses as well.

We first focus on a single snapshot at a certain clock cycle leading to binary observations denoted by $\{\omega_0,\ldots,\omega_{n-1}{\in\mathbb{F}_2}\}$ corresponding to $n$ registers of the design.
Some registers belong to the control logic (finite-state machine), which are out of our interest.
Therefore, we target $m \leq n$ registers according to the architecture of the underlying design.
For example, $m=256$ for an unprotected implementation of AES (128 bits for the state register and 128 bits for the key register).
If we define variables $v_{i\in\{0,\ldots,m-1\}}$ for the value of targeted register cells at the selected clock cycle, we can write
\begin{equation}
v_i=c^i_0\omega_0 + ... + c^i_{n-1}\omega_{n-1},
\label{eq:app:var}
\end{equation}
where with $c^i_j$ we denote binary coefficients.
Since only one of the observations is associated to the $i$-th register cell, only one of the coefficients $c^i_{j\in\{0,\ldots,n-1\}}$ is 1, and the rest are 0.
In other words, $\forall i, \sum\limits_{\forall j}{c^i_j}=1$.
These are the first formulations that we require to include in the Boolean satisfiability problem~(SAT).
To this end, we break the addition into bit level by defining intermediate variables $t_{j\in\{2,\ldots,n-1\}}$ for each $i$ individually.
Below, we drop the superscript $i$ for both $t$ and $c$ for simplicity.
Adding $c_0$ and $c_1$ leads to result $t_2=c_0 \oplus c_1$ and carry $c_0c_1$.
Since the carry must be zero, we can add the following clauses to the SAT.
\begin{equation}
\overline{t_2} \oplus c_0 \oplus c_1=1, ~~~~~~ \overline{c_0}\,\lor\,\overline{c_1}=1
\label{eq:app:HW1}
\end{equation}
The same procedure is repeated for adding $c_2$ and the result of former addition $t_2$, i.e., $t_3=c_2 \oplus t_2$ and $c_2t_2=0$.
Generally, we can write
\begin{equation}
\forall j\in\{2,\ldots,n-2\}, ~~~ \overline{t_{j+1}} \oplus c_j \oplus t_j =1,~~~~~ \overline{c_j}\,\lor\,\overline{t_j}=1
\label{eq:app:HW2}
\end{equation}
At the end, we add a clause $t_{n-1} \oplus c_{n-1}=1$ to the SAT, defining that the final result of the addition should be 1.
These clauses (which are independent of the observations $\omega$ and the architecture of the circuit under attack) are generated individually for each targeted register cell $i\in\{0,\ldots,m-1\}$.

We should also add the CNF of~\eqref{eq:app:var} for each targeted register cell.
Those observations $\omega_j$ that are 0 cancel out the corresponding coefficient $c_j$.
Therefore, we can write
\[
v_i \oplus \biggl(\sumsc\limits_{\forall j, \omega_j=1}{c_j}\biggr)=0.
\]
This translates to 
\begin{equation}
\overline{v_i} \lor \left(\lorrsc\limits_{\forall j, \omega_j=1}{c_j}\right)=1, ~~~~~~
v_i \lor \overline{\left(\lorrsc\limits_{\forall j, \omega_j=1}{c_j}\right)}=1.
\label{eq:app:var1}
\end{equation}
The left equation can be easily added as a clause to the SAT (as it is already in CNF), but the right one should be split into multiple clauses as follows:
\begin{equation}
\forall j, \omega_j=1, ~~~~~ v_i \lor \overline{c_j}=1.
\label{eq:app:var2}
\end{equation}

Having more snapshots at different clock cycles, the clauses in~\eqref{eq:app:var1} and~\eqref{eq:app:var2} should be repeated for $m$ distinct register variables $v_i$ based on the corresponding observations $\omega_j$.
However, the coefficients $c^i_j$ stay the same, i.e., they are defined only once for the entire circuit independent of the number of snapshots.
Accordingly, the clauses in~\eqref{eq:app:HW1} and~\eqref{eq:app:HW2} are also not repeated.

The remaining task is to link the variables $v_i$ (of targeted register cells) at different clock cycles.
This is done based on the underlying design architecture of the circuit under attack and the functions it realizes.
For example, in a round-based architecture, the state register cells store the output of the cipher round function, and the key register cells the round keys.
In a serialized architecture, the content of the registers is shifted (e.g., in a byte-wise fashion), and certain operations (e.g., Sbox) are applied on particular registers at determined clock cycles.

In case of a masked implementation with $d+1$ shares, the number of targeted registers at each clock cycle becomes $m(d+1)$ (for example, $512 \times 2$ for a first-order masked implementation of AES making use of the state and key registers with 2 shares).
Therefore, the entire clauses given in~\eqref{eq:app:var1} to~\eqref{eq:app:var2} should be repeated $d+1$ times.
In the next step, we define $m$ virtual variables $\nu_i= \bigoplus\limits_{l=1}^{d+1}v_{i,l}$ (for each clock cycle), where $(v_{i,1}, \ldots, v_{i,d+1})$ represent variable $\nu_i$ with $d+1$ shares.
The corresponding clauses can be written as
\[
\forall i\in\{1,\ldots,m\}, ~~~~~ \overline{\nu_i} \oplus v_{i,1} \oplus \ldots \oplus v_{i,d+1}=1.
\]
The rest is similar to an unmasked implementation, i.e., the (unmasked) variables $\nu_i$ at different clock cycles are linked based on the design architecture of the circuit under attack.

\end{document}